\documentclass[useAMS,usenatbib]{mn2e}
\usepackage{url,times,graphicx,amsmath,amsfonts,amssymb,aas_macros,color,epsfig,varioref,subfigure}

\newcommand{\Tab}[1]{Table~\ref{#1}}
\newcommand{\Sec}[1]{Section~\ref{#1}}
\newcommand{\Eq}[1]{Eq.~(\ref{#1})}
\newcommand{\Fig}[1]{Fig.~\ref{#1}}
\newcommand{\hMpc}{{\ifmmode{h^{-1}{\rm Mpc}}\else{$h^{-1}$Mpc}\fi}}
\newcommand{\hGpc}{{\ifmmode{h^{-1}{\rm Mpc}}\else{$h^{-1}$Gpc}\fi}}
\newcommand{\hkpc}{{\ifmmode{h^{-1}{\rm kpc}}\else{$h^{-1}$kpc}\fi}}
\newcommand{\hMsun}{{\ifmmode{h^{-1}{\rm {M_{\odot}}}}\else{$h^{-1}{\rm{M_{\odot}}}$}\fi}}
\def\hMpc{$h^{-1}\,{\rm Mpc}$}
\def\hkpc{$h^{-1}\,{\rm kpc}$}
\def\LCDM{\ensuremath{\Lambda}CDM}
\def\ude{uDE}
\def\cde{cDE}

\title
[Halo properties and the cosmic web]
{Hydrodynamical simulations of coupled and uncoupled quintessence models I:  
Halo properties and the cosmic web}
\author[Carlesi et al.]
{Edoardo Carlesi,$^{1}$
\thanks{E-mail: edoardo.carlesi@uam.es}
Alexander Knebe,$^{1}$ Geraint F. Lewis,$^{2}$ Scott Wales,$^{2,3}$ Gustavo Yepes$^{1}$
\\
$^{1}$Departamento de F\'isica Te\'orica,
Universidad Aut\'onoma de Madrid, 28049, Cantoblanco, Madrid, Spain\\
$^{2}$Sydney Institute for Astronomy, School of Physics, A28, 
The University of Sydney, NSW 2006, Australia\\
$^{3}$ARC Centre of Excellence for Climate System Science, 
School of Earth Sciences, The University of Melbourne, Australia 3010}
\begin{document}

\date{Accepted XXXX . Received XXXX; in original form XXXX}

\pagerange{\pageref{firstpage}--\pageref{lastpage}} \pubyear{2013}

\maketitle

\label{firstpage}

\begin{abstract}
  We present the results of a series of adiabatic hydrodynamical
  simulations of several quintessence models (both with a free and an
  interacting scalar field) in comparison to a standard \LCDM\
  cosmology.  For each we use $2\times1024^3$ particles in a
  $250$\hMpc\ periodic box assuming WMAP7 cosmology.  In this work we
  focus on the properties of haloes in the cosmic web at $z=0$.  The
  web is classified into \emph{voids}, \emph{sheets}, \emph{filaments}
  and \emph{knots} depending on the eigenvalues of the velocity shear
  tensor, which are an excellent proxy for the underlying overdensity
  distribution.  We find that the 
  properties of objects classified according to their surrounding environment
  shows a substantial dependence on the underlying cosmology;
  for example, while
  $V_{\rm max}$ shows average deviations of $\approx5$ per cent across
  the different models when considering the full halo sample, 
  comparing objects classified according to
  their environment, the size of the deviation can be as large as $20$
  per cent.

  We also find that halo spin parameters are positively correlated to
  the coupling, whereas halo concentrations show the opposite
  behaviour. Furthermore, when studying the concentration-mass relation in
  different environments, we find that in all cosmologies underdense
  regions have a larger normalization and a shallower slope.  While
  this behaviour is found to characterize all the models, differences
  in the best-fit relations are enhanced in (coupled) dark energy
  models, thus providing a clearer prediction for this class of
  models.

\end{abstract}

\begin{keywords}
methods:$N$-body simulations -- galaxies: haloes -- cosmology: theory -- dark matter
\end{keywords}

\section{Introduction}
Over more than 15 years, since observations of high-redshift Supernovae of type Ia \citep[see][]{Riess:1999, Perlmutter:1999} first
indicated  that the Universe is undergoing an accelerated expansion, a large number of cosmological probes, including
cosmic microwave background (CMB) anisotropies \citep[][]{Wmap:2011, Sherwin:2011}, weak lensing 
\citep{Huterer:2010}, baryon acoustic oscillations (BAO) \citep{Beutler:2011} and large scale structure (LSS) surveys
\citep{SDSS:2009}, have confirmed this startling claim and shown that the Universe is spatially flat.
To explain these diverse observation, cosmology requires the presence
of a fluid, called dark energy (DE), which permeates the whole Universe and 
exerts a negative pressure, eventually overcoming the gravitational pull that would otherwise dominate.
The standard model of cosmology, referred to as \LCDM, provides the simplest
possible explanation for DE, assuming that DE is played by a constant called $\Lambda$
which possesses a constant equation of state, such that $p_{\Lambda}=-\rho_{\Lambda}$.  
However, despite its simplicity and observational viability, \LCDM\ still lacks of appeal
from a purely theoretical point of view, due to \emph{fine tuning} and 
\emph{coincidence} problems;
the first refers to the fact that, if we assume that $\Lambda$ is
the zero-point energy of a fundamental quantum field, to be compatible
with the aforementioned cosmological constraints its density requires an unnatural
fine-tuning of several tens of orders of magnitude.
The second problem arises from the difficulty in explaining in a satisfactory way 
the fact that matter and dark energy densities \emph{today} have comparable values, 
although throughout most of the cosmic history their evolutions have followed completely
different patterns.

It is thus natural to explore the possibility that dark energy does
not take the form of a cosmological constant, $\Lambda$, but is instead
a dynamical component of the universe, whose energy density evolves
with time, eventually dominating in the present epoch.  In this sense,
a large number of different models, such as Chaplygin gas
\citep{Kamenshchik:2001}, vector dark energy
\citep[][]{BeltranMaroto:2008, Carlesi:2012}, $\kappa$-essence
\citep{Picon:2000} and quintessence \citep{Wetterich:1995,
Caldwell:1998, Copeland:1998, Zlatev:1999}
have been proposed to overcome the perceived theoretical
shortcomings of the standard cosmology.  In particular, quintessence
(or scalar field) models are viable and likely candidates for
dynamical dark energy \citep[see][]{Tsujikawa:2013}, as they can
reproduce current observational data without being plagued by the
fine-tuning problem of \LCDM, since their expansion history - at least
for a set of different potentials - is almost insensitive to the
particular choice of the field's initial conditions.  An interesting
subset of quintessence theories is represented by coupled models,
where it is assumed that the scalar field has a non-negligible
interaction to the dark matter sector \citep{Amendola:2000} and is
thus expected to leave a strong imprint on structure formation.

While both classes of quintessence models have been already studied
numerically by means of $N$-body simulations (see for instance
\cite{Klypin:2003}, \cite{DeBoni:2011} for free and
\cite{Baldi:2010a}, \cite{Li:2011} for coupled models), in the present
work we aim at investigating and highlighting the differences arising among
coupled and uncoupled scalar fields with the same potential, and
compare our results to a benchmark \LCDM\ cosmology.  
Our aim is to disentangle the effects due to the \emph{fifth force}
acting on dark matter particles from those caused to the dynamical
nature of dark energy, when considering the deeply non-linear regime of
the models. This way we will discern strategies to
observationally distinguish between coupled and uncoupled forms of
quintessence and thus to provide new tools for model selection.  

Using a suitably
modified version of the publicly available SPH/$N$-body code
\texttt{GADGET-2} \citep{Springel:2005} we undertake a series of simulations
of different quintessence models with a Ratra-Peebles \citep{Ratra:1988} potential and
several values of the coupling parameter allowed by current
observational constraints.
The box size ($250$\hMpc), the number of particles ($2\times1024^3$) and
the use of adiabatic smoothed particle hydrodynamics allow us to
analyse a large amount of different properties with a good resolution
and statistics.  In this first of a series of papers, we will consider large-scale structures (LSS) and its environment, with the physics of galaxy clusters presented
in a follow-up paper (Paper II).
In the present work,
we analyze in particular the structure of the cosmic web and the correlations between the
environment, dark matter haloes and gas across these different
cosmological models.

The paper is organized as follows; in \Sec{sec:model}, we briefly recall the general features
of the quintessence models considered in this work as well as of the recipes 
necessary for their simulation using $N$-body techniques. 
In \Sec{sec:setup}, we discuss the
settings of our particular simulations as well as those of the halo
finder, together with the classification of the cosmic web.
In \Sec{sec:halo_res} we present LSS properties of the
modified frameworks, \Sec{sec:web_res} is dedicated to the general
features of the cosmic web while in \Sec{sec:haloweb_res} we describe
the results of the correlation of haloes to their environment.  A
summary of the results obtained and a discussion on their
implications is then presented in section \Sec{sec:conclusions}.

\section{Prerequisites}\label{sec:model}
Here, we will briefly recall the basic properties of quintessence models and their
implementation into a cosmological $N$-body algorithm.
We refer the reader to the works of \cite{Wetterich:1995, Amendola:2000, Amendola:2003, Pettorino:2012, Chiba:2013}
for discussions on the theoretical and observational properties of (coupled) quintessence models, 
and to \cite{Maccio:2004, Baldi:2010a, Li:2011d} for a thorough description of the numerical approaches.

\subsection{The models}
In quintessence models the role of dark energy is played by a cosmological 
scalar field $\phi$ whose Lagrangian can be generally written as:
\begin{equation}\label{eq:lagrangian}
L = \int d^4x \sqrt{-g} \left(-\frac{1}{2}\partial_{\mu}\partial^{\mu}\phi 
	+ V(\phi)+ m(\phi)\psi_{m}\bar{\psi}_{m} \right) 
\end{equation}
where in principle $\phi$ can interact with the dark matter field $\psi_m$ through its mass term, meaning that,
in general, dark matter particles will have a time-varying mass.
With a suitable choice of the potential $V(\phi)$, quintessence cosmologies can account 
for the late time accelerated expansion of the universe both in the interacting and non interacting case. 
In the present work we have focused on the so called Ratra-Peebles
\citep[see][]{Ratra:1988} self interaction potential: 
\begin{equation}\label{eq:ratra}
V(\phi) = V_0\left(\frac{\phi}{M_p}\right)^{-\alpha}
\end{equation}
where $M_p$ is the Planck mass, while $V_0$ and $\alpha$ 
are two constants whose values can be fixed by fitting the model to observational data
\citep[see][]{Wang:2012, Chiba:2013}.

In \Eq{eq:lagrangian}, we allowed for the scalar field to interact with matter through the mass term 
$m(\phi)\psi\bar{\psi}$; a popular choice \citep[see]{Pettorino:2012} for the function $m(\phi)$ is:
\begin{equation}\label{eq:mass}
m(\phi) = m_0 \exp{\left(-\beta(\phi)\frac{\phi}{M_p}\right)} 
\end{equation}
which is also the one assumed in this \emph{paper}.

In the following, we have taken into account a constant interaction term $\beta(\phi)=\beta_0$,
which from \Eq{eq:mass} implies an energy flow from the dark matter to the dark energy sector and thus a diminishing mass
for dark matter particles.
\begin{table}
\caption{Values of the coupling and potential used for the \ude and \cde\ models.}
\label{tab:parameters}
\begin{center}
\begin{tabular}{cccc}

\hline
Model		& $V_0$		& $\alpha$ 	&  $\beta$ \\
\hline
\ude 	 	& $10^{-7}$ 	& $0.143$	& $-$ \\
\cde033		& $10^{-7}$ 	& $0.143$ 	& $0.033$\\
\cde066		& $10^{-7}$ 	& $0.143$ 	& $0.066$\\
\cde099		& $10^{-7}$ 	& $0.143$ 	& $0.099$\\
\hline
\end{tabular}
\end{center}
\end{table}
In \Tab{tab:parameters} we list the values for $V_0$, $\alpha$ and
$\beta$ of \Eq{eq:ratra} and \Eq{eq:mass} as used in the four
non-standard cosmologies under investigation - an uncoupled Dark
Energy (\ude) model and three coupled Dark Energy (\cde) ones. The
latter differ only by the choice of the coupling and have been named
accordingly. The particular values used in all the implementations
have been selected according to the Cosmic Microwave Background (CMB) 
constraints discussed in \cite{Pettorino:2012}, to ensure the cosmologies under
investigation to be compatible with the WMAP7 dataset \citep{wmap7}. 
However, more recent results obtained using Planck data \citep[see][]{Pettorino:2013}, 
provide even tighter constraints on the free parameters of these models, which
shall be the object of subsequent investigation.

\subsection{Numerical implementation}\label{sec:code}
\begin{table}
\caption{Cosmological parameters at $z=0$ used in the \LCDM, \ude, \cde033, \cde066 and \cde099 simulations.}
\label{tab:cosmology}
\begin{center}
\begin{tabular}{cc}
\hline
Parameter & Value \\
\hline
$h$		& $0.7$   \\
$n$		& $0.951$   \\
$\Omega_{dm}$  	& $0.224$ \\
$\Omega_b$   	& $0.046$ \\
$\sigma_8$   	& $0.8$   \\
\hline
\end{tabular}
\end{center}
\end{table}
The first simulations of interacting dark energy models were performed by \cite{Maccio:2004}, who described
the basic steps for implementing interacting quintessence into the \texttt{ART} code. 
In our case, we built our implementation on \texttt{P-GADGET2}, a modified version
of the publicly available code \texttt{GADGET2}
\citep[][]{Springel:2001, Springel:2005}. This version has first been
developed to simulate vector dark energy models
\citep[see][]{Carlesi:2011, Carlesi:2012} and was then extended to
generic dynamical dark energy as well as coupled dark energy
cosmologies. The algorithm used is based on the standard Tree-PM
solver with some modifications added to take into account the
additional long-range interactions due to the coupled scalar field
which effectively act as a rescaling of the gravitational constant. 
For the implementation of these features non-standard models we followed
closely the recipe described in \cite{Baldi:2010a}, to which the reader is referred.

This approach requires that a number of quantities, namely:
\begin{itemize} 
\item the full evolution of the scalar field $\phi$ and its derivative $\dot{\phi}$,
\item the variation mass of cold dark matter particles $\Delta m(z)$, and
\item the background expansion $H(z)$. 
\end{itemize}
have to be computed in advance and then interpolated at run time.
We therefore implemented background and first order Newtonian
perturbation equations into the publicly available Boltzmann code
\texttt{CMBEASY} \citep{Doran:2005} to generate the tables containing
the aforementioned quantities. 
The starting background densities
were chosen in order to ensure the same values at $z=0$ for the
cosmological parameters listed in \Tab{tab:cosmology}; linear
perturbations have been solved assuming adiabatic initial conditions.

Finally, in the case of non-standard cosmologies it is necessary to properly
generate the initial conditions of the $N$-body simulations taking
into account not only the different matter power spectra but also
the altered growth factors and logarithmic growth rates, respectively.
These are in fact the necessary ingredients to compute the initial
particles' displacements and velocities on a uniform Cartesian grid
using the first order Zel'dovich approximation \citep{Zeldovich:1970}.
We implemented these changes into the publicly available
\texttt{N-Genic}\footnote{http://www.mpa-garching.mpg.de/gadget/} MPI
code, which is suitable for generating \texttt{GADGET} format initial
conditions.  Again, the matter power spectra, growth
factors $D(t)$ and logarithmic growth rates $f=d \ln D(t) / d \ln a$
  have been computed for the four non-standard cosmologies using the
  modified \texttt{CMBEASY} package.

All the above changes have been carefully tested against theoretical
predictions and the previous results existing in the literature to ensure
the consistency and reliability of our modifications.

\section{The simulations}\label{sec:setup}
\subsection{Settings}
\begin{table}
\caption{$N$-body settings and cosmological parameters used for the three simulations.}
\begin{center}
\begin{tabular}{cc}
\hline
Parameter & Value \\
\hline
$L_{box}$	& $250$\hMpc \\
$N_{dm}$ 	& $1024^3$ \\
$N_{gas}$	& $1024^3$ \\
$m_{dm}$	& $9.04 \times 10^8$\hMsun \\
$m_{b}$		& $1.85 \times 10^8$\hMsun \\
$z_{start}$	& $60$ \\
\hline
\end{tabular}
\label{tab:settings}
\end{center}
\end{table}
Our set of $N$-body simulation has been devised in order to allow us
to compare and quantitatively study the peculiarities of the different
models in the physics of galaxy clusters and the properties of the
cosmic web.  To do this, we have chosen a box of side length
$250$\hMpc\ (comoving) where we expect to be able to analyze with adequate
resolution a statistically significant ($>100$) number of galaxy
clusters ($M>10^{14}$\hMsun) as well as the properties of the
different cosmic environments, classified as voids, sheets, filaments
and knots.  The parameters chosen to set up the simulations, which are
common to all the six models under investigation, are listed in
\Tab{tab:settings}.

In this series of simulations we implemented adiabatic SPH
only, thus neglecting the effects of all sources of radiative effects \citep[][]{Monaghan:1992, Springel:2010}
This way we are able to establish a clear basis for the
differences induced on baryons by the different cosmologies, without
the need to take into account the additional layer of complexity
introduced of radiative physics, which in itself requires a
substantial degree of modeling.
The publicly available version of \texttt{GADGET-2} performs a 
Lagrangian sampling of the continuous fluid quantities using a set 
of discrete tracer particles. Gas dynamics equations are then solved
using the SPH entropy conservation scheme described in \cite{Springel:2005}.
In our case, continuous fluid quantities are computed using a number of smoothing 
neighbours $N_{sph}=40$.
Gas pressure and density are related through the relation $P\propto\rho^{\gamma}$,
where $\gamma=\frac{5}{3}$ under the adiabatic assumption.
 
In addition to the four quintessence models (whose parameters have
been given in \Tab{tab:parameters}) we simulated a \LCDM\ cosmology,
which we use as a benchmark to pinpoint deviations from the standard
paradigm.  The initial conditions for all the simulations have been
generated using the same random phase realization for the Gaussian
fluctuations, which enables us
to consistently cross-correlate properties enforcing the same values at present time for
$\Omega_m$ and $\sigma_8$ (cf. \Tab{tab:cosmology}), 
across different simulations.

\subsection{Halo identification}
We identified haloes in our simulation using the open
source halo finder
\texttt{AHF}\footnote{\texttt{AHF} stands for
  \texttt{A}miga~\texttt{H}alo~\texttt{F}inder, which can be downloaded
  freely from \texttt{http://www.popia.ft.uam.es/AHF}} 
described in \cite{Knollmann:2009}; this code improves the
\texttt{MHF} halo finder \citep{Gill:2004} and has been widely
compared to a large number of alternative halo finding methods \citep{Knebe:2011,Onions:2012,Knebe:2013}.  
\texttt{AHF} computes the density field and locates the 
prospective halo centres at the local overdensities.
For each of these density peaks, it determines the gravitationally
bound particles, retaining only peaks with at least 20 of them,
which are then considered as haloes and further analyzed.

The mass is computed via the equation
\begin{equation} \label{eq:virial_mass_definition}
M_{\Delta} = \Delta \times \rho_{c}(z) \times \frac{4 \pi}{3} R_{\Delta}^{3}.
\end{equation}
so that $M(R)$ is defined as the total mass contained within a
radius $R$ at which the halo matter overdensity
reaches $\Delta$ times the critical value $\rho_c$. 
Since the critical density of the universe is a function of redshift,
we must be careful when considering its definition, which reads
\begin{equation}
\rho_{c}(z) = \frac{3 H^2(z)}{8 \pi G} 
\end{equation}
as the evolution of the Hubble parameter, $H(z)$ differs at all redshifts in the five models.  
In the latest version of
\texttt{AHF} this problem is solved reading the $H(z)$ for the \cde\ and \ude\ models
in from a precomputed table, which then allows to compute the $\rho_c(z)$ consistently in each case.
For all models we assume $\Delta=200$.

\subsection{Classification of the cosmic web}\label{sec:web}
As we intend to correlate halo properties with the environment, it is
necessary to introduce the algorithm used for the classification of
the cosmic web into voids, sheets, filaments and knots.  Using the
term cosmic web \citep{Bond:1996} we refer to the complex visual
appearance of the large scale structure of the universe, characterized
by thin linear filaments and compact knots crossing regions of very
low density \citep{Massey:2007, Kitaura:2009, Jasche:2010}.

The exact mathematical formulation for describing the visual
impression of the web is highly non-trivial and can be implemented
using two different approaches, the geometric one and the dynamic one.
The first one relies on the spatial distribution of haloes in
simulations \citep{Novikov:2006, Aragon-Calvo:2007} disregarding the
dynamical context. The second approach starts with the classification
of \cite{Hahn:2007}, where they identified the type of environment
using the eigenvalues of the tidal tensor (i.e. the Hessian of the
gravitational potential), rather than studying the matter density
distribution.

However, these particular approaches are unable to resolve the web on
scales smaller than a few megaparsecs \citep{Forero-Romero:2009}. While
retaining the original idea of dynamical classification,
\cite{Hoffman:2012} proposed to replace the tidal tensor with the
velocity shear, showing that this approach has a much finer resolution
on the smaller scales while reproducing the large scale results of the
other approach. Defining the velocity shear tensor as
\begin{equation}
\Sigma_{\alpha\beta} = -\frac{1}{2 H_0}\left(\frac{\partial v_{\alpha}}{\partial r_{\beta}} - \frac{\partial v_{\beta}}{\partial r_{\alpha}} \right)
\end{equation}
and diagonalizing it, we obtain the eigenvalues $\lambda_1, \lambda_2$
and $\lambda_3$. Taking the trace of $\Sigma_{\alpha\beta}$ we obtain
\begin{equation}
Tr(\Sigma_{\alpha\beta}) = \sum_i \lambda _i = -\overrightarrow{\nabla}\cdot\overrightarrow{V}\propto \delta_m 
\end{equation}
from which we see that there is indeed a direct relationship between the eigenvalues of the 
velocity shear tensor and the matter overdensity. In practice the eigenvalue $\lambda _i$
is related to the intensity of the inflow (outflow) of matter along the $i$-th axis in the base 
where $\Sigma_{\alpha\beta}$ is diagonal.

We therefore proceed to classify the cosmic web ordering the
eigenvalues $\lambda _1 > \lambda _2 > \lambda_3$ and defining the
different points on the web as
\citep{Hoffman:2012,Libeskind:2012,Libeskind:2013}:
\begin{itemize}
\item \emph{voids}, if $\lambda _1 < \lambda _{th}$ \\
\item \emph{sheets}, if $\lambda _1 > \lambda_{th} > \lambda _2$ \\
\item \emph{filaments}, if $\lambda _2 > \lambda_{th} > \lambda_3$ \\
\item \emph{knots}, if $\lambda _3 > \lambda_{th}$
\end{itemize}
where $\lambda_{th}$ is a free threshold parameter (to be specified below).

The computation of the eigenvalues has been performed on a regular
$256^3$ grid, corresponding to a cell size of $0.97$\hMpc. We use a
triangular-shaped cloud (TSC) prescription for the assignment of the particles
\citep{Hockney:1988} and then compute the overdensity and the
eigenvalues of the velocity shear tensor for every grid cell.  Using
the \texttt{AHF} catalogues, we assign every halo to the nearest grid
point hence providing us with a measure of environment for every
object.

At this stage we still have not explicitly
classified the cosmic web, as we lack a clear theoretical prescription
for the value of $\lambda_{th}$.  In our case, we have fixed
$\lambda_{th}$ to the highest value which ensures that no halo with
$M>10^{14}$\hMsun\ belongs to a void in any simulation.  At a first
glance, this kind of constraint might seem redundant, as it would be
implied in any standard definition of void as an underdense region.
However, we must recall here that our definition of the cosmic web
relies solely on the dynamical properties of the matter distribution
(being related to the magnitude of its inflow or outflow in a given
node) and may in principle overlook its net density content.  It is
thus necessary to enforce this principle explicitly tuning our free
parameter to $\lambda_{th}=0.1$, which is the value which in this case
satisfies the aforementioned condition and has been used in
\Sec{sec:web_res}.  For a more elaborate discussion of $\lambda_{th}$
we refer the reader to \citet{Hoffman:2012}; we only note that our
choice is close to their proposed value.

\section{Large scale clustering and general properties}\label{sec:halo_res}
Before presenting the results relative to the properties of the cosmic
web and the correlation of halo properties to the environment, we will
describe some aspects of large scale structure (LSS) and general halo
properties in our simulations. This should give a more traditional
overview of the effects of (coupled) dark energy models.

\subsection{Halo mass function}
The halo mass function (HMF) in coupled dark
energy cosmologies has already been studied by \cite{Maccio:2004, Nusser:2004, Baldi:2010a, 
Li:2011, Cui:2012} so that we will only briefly comment on the
topic. Our results reproduce the earlier findings of 
Baldi's  in the overlapping regions of mass and $k$-space,
thus providing an additional proof of the correct functioning of our
modified implementation.
\begin{figure*}
\begin{center}
\begin{tabular}{ccc}
\includegraphics[height=8.0cm]{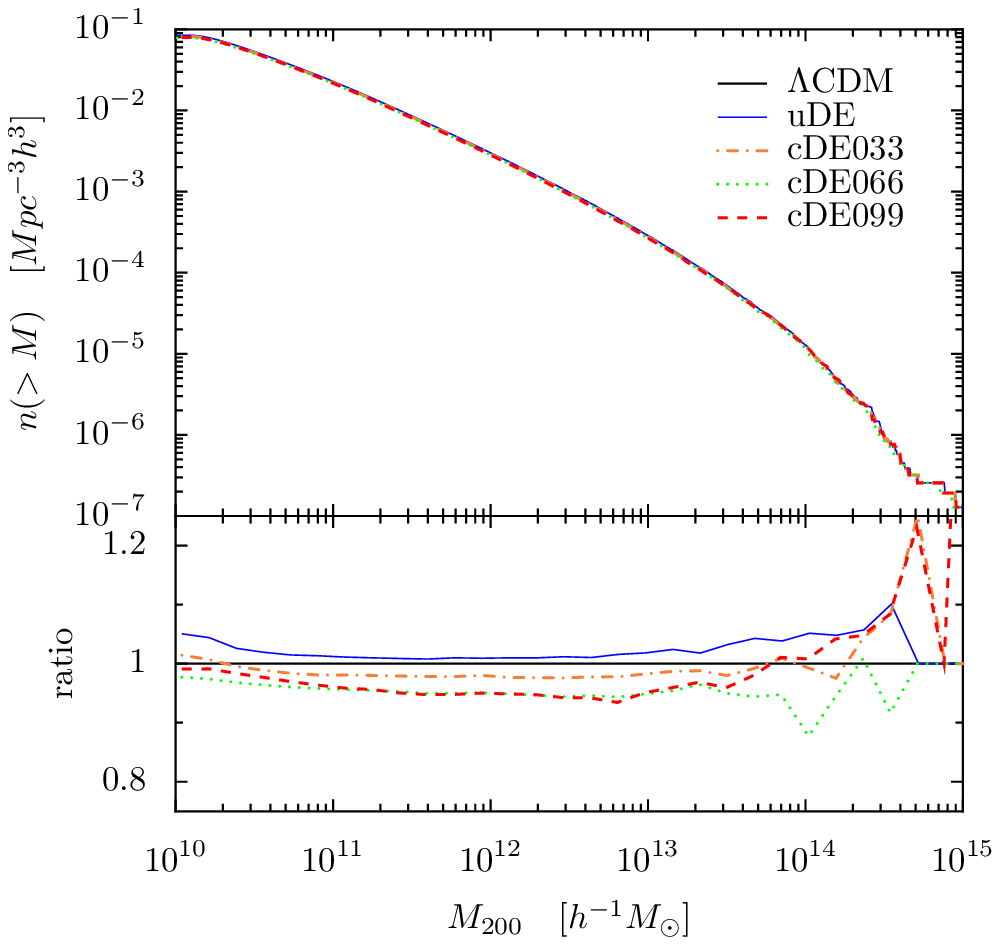} &
\includegraphics[height=8.0cm]{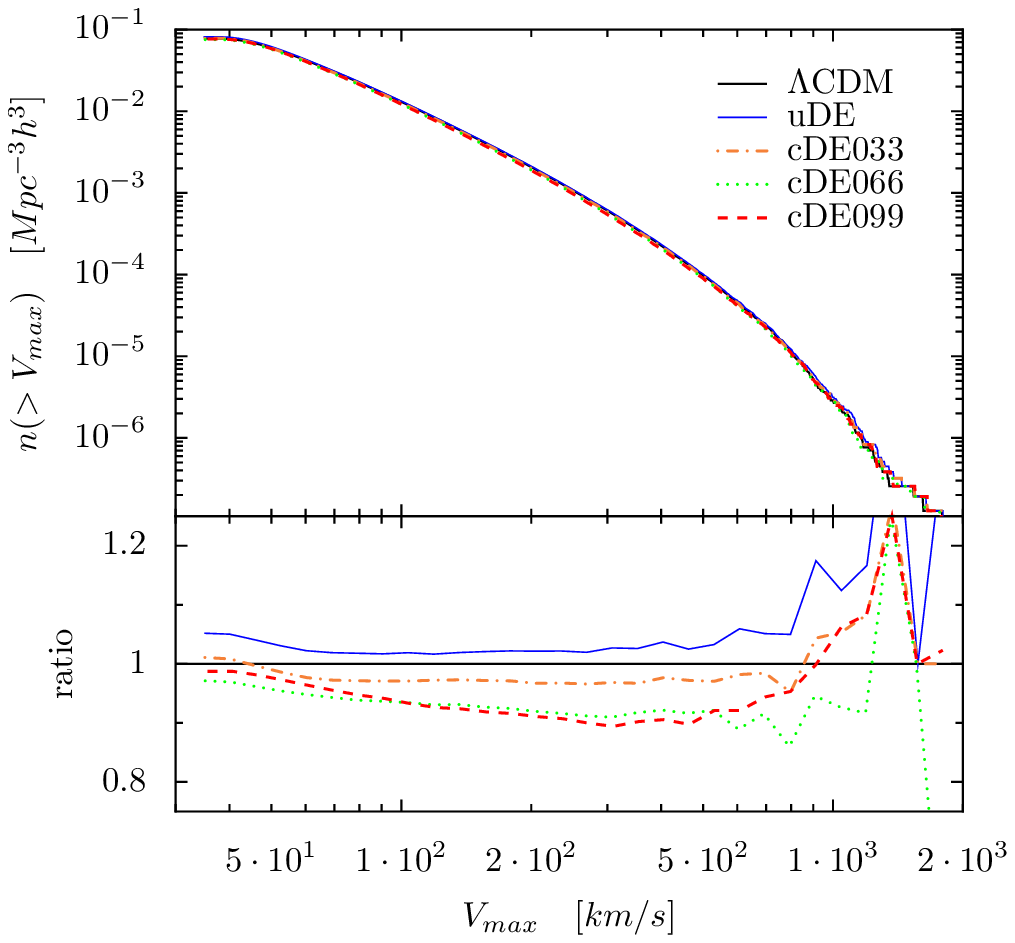} \\
\end{tabular}
\caption{\small Cumulative halo density number counts as a function of mass (left panel)
and $V_{max}$ (right panel). Compared to \LCDM, a slight suppression of the number of objects produced at redshift $z=0$
is predicted for \cde\ models while the opposite is true for \ude.
Albeit small, the effect is enhanced in the velocity function, where the strongest coupled model
differs up to $10\%$ from \LCDM\ (neglecting the higher mass ends, which are affected
by a very low statistics).}
\label{img:numcount}
\end{center}
\end{figure*}

In \Fig{img:numcount} we show the cumulative mass (left) and velocity
functions (right) as well as the ratio to \LCDM\ for the four
quintessence models.  Singling out the region from $10^{10}$ to
$10^{14}$ \hMsun, and neglecting the higher mass end, where the
statistics is unreliable due to the low number of objects, we can
see that the largest difference in number counts amounts to $\approx
7\%$ for the strongest coupled models, gradually decreasing for
smaller couplings.  In the velocity function, this suppression reaches
$10\%$, thus slightly enhancing the magnitude of the effect.

We can compare our results for the HMF with those of \cite{Cui:2012},
who modeled the \cite{Jenkins:2001} and \cite{Tinker:2008} mass functions for a series of similar coupled
dark energy models, using Friends-of-Friends (FoF) and Spherical
Overdensity (SO) algorithms to build up their halo catalogues.  Even
though in their simulations they used different $\sigma_8$
normalizations, fitting the analytical HMFs to the numerical results
they were able to extend the predictions for \cde\ cosmology to
arbitrary $\sigma_8$ values. Using the same \LCDM\ $\sigma_8$
normalization, then, they also found a $5-10\%$ suppression of the HMF
of \cde, in perfect agreement with our results.

Although not shown here, we have also verified that these $z=0$
results match the analytical prediction of the Tinker mass function
\citep[]{Tinker:2008}, provided the correct input power spectra and
normalizations are used.  We can safely conclude that the presence of
coupled and uncoupled quintessence of the kind described here is
expected to produce differences from \LCDM\ predictions up to a factor
$10\%$ in present day's HMF.  Remarkably, this estimate is
qualitatively independent of the algorithm used for the halo
identification as we have seen comparing our results to the work of
\cite{Cui:2012}.

\begin{table*}
  \caption{
    Total number of haloes found in each simulation corresponding to
    our applied mass cuts of $M>3\times10^{11}$\hMsun\ and the
    $M>10^{12}$\hMsun, respectively. }
\label{tab:n_haloes}
\begin{center}
\begin{tabular}{cccccc}
\hline
Mass cut 		& \LCDM	 & \ude 	  & \cde033    & \cde066 & \cde099 \\
\hline 
$M>3\times10^{11}$\hMsun &  $138211$ & $139288$ &  $135613$  &  $130877$  & $130812$\\
$M>10^{12}$\hMsun 	 &  $46196$ & $46179$ &  $44943$  &  $44363$  & $43749$\\
\hline
\end{tabular}
\end{center}
\end{table*}

\subsection{Halo properties}\label{sec:halop}
To study internal halo properties (such as spin parameter and
concentration) we first need to define a statistically sound sample of
objects, in order to reduce the impact of spurious effects on the
results.  This means that we need to constrain our analysis to
structures which satisfy some conditions on both \emph{resolution} and
\emph{relaxation}.

The first condition means that we have to restrict our analysis to
objects with a number of particles above a given threshold, taking
into account the existing trade-off between the quality and the size
of the halo sample. The second criterion needs to be applied as we
want to focus on structures as close as possible to a state of
dynamical equilibrium.  In fact, many phenomena, such as infalling matter
and major mergers, may take place, driving the structure out of 
equilibrium. In this case, then, the determination of quantities such as
density profiles and concentrations becomes unreliable 
(see for instance \citet{Maccio:2007} and \cite{Munoz-Cuartas:2011}).

Following \cite{Prada:2012}, we will define as relaxed only
the haloes that obey to the condition
\begin{equation}
\frac{2K}{|U|} - 1 < 0.5
\end{equation}
without introducing other selection parameters; for alternative
ways of identifying unrelaxed objects we refer for instance to
\citet{Maccio:2007, Bett:2007, Neto:2007,Knebe:2008, Prada:2012, Munoz-Cuartas:2011,Power:2011}. 
For the moment, we neglect the impact of \ude\ and \cde\ on the
definition of the virial ratio since this effect is of just a few
percent \citep{Abdalla:2010, Pace:2010} and is thus subleading in our
case, where we are removing objects off by more than $50\%$ from the
standard relation.

Now that we have established the rules that will shape our halo sample, we proceed
to study  some internal properties of dark matter haloes,
namely, spin and concentration -- as a function of halo mass --
enforcing one additional criterion for the halo selection: the number
of particles in it. When studying the spin parameter, we will restrict
ourselves to haloes with $M_{200} > 3\times10^{11}$\hMsun,
i.e. composed of at least $\approx 600$ baryon and dark matter
particles, following the choices of \cite{Bett:2007},
\cite{Maccio:2007}, \cite{Munoz-Cuartas:2011} and \cite{Prada:2012}.
In the case of halo concentrations, we applied a stricter criterion,
using $M_{200} > 1\times10^{12}$\hMsun (or $\approx 2000$ particles),
due to the fact that the computation of halo concentration requires a
better resolution of the central regions, as we will discuss in the
dedicated subsection.

\begin{figure}
\begin{center}
\includegraphics[width=8cm]{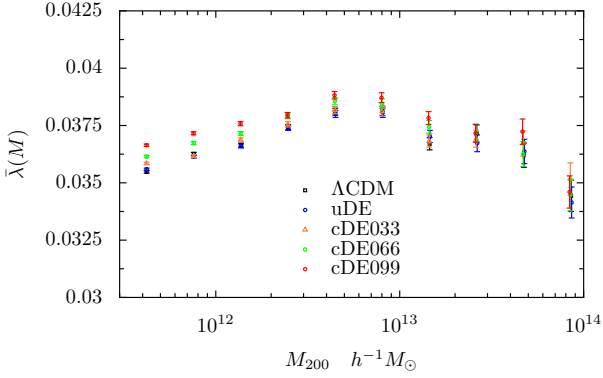}
\caption{\small Average value of the spin parameter per mass bin.
We can see that the
spin parameter has a weak positive correlation to the mass until $\approx 8\times10^{12}$\hMsun
and a negative one after that threshold. 
Further, haloes in coupled dark energy models have an average value which is
slightly larger than uncoupled ones.}
\label{img:spin}
\end{center}
\end{figure}

\subsubsection{Spin parameter}
We can study the rotational properties of haloes introducing the so-called spin parameter $\lambda$
\citep[e.g.][]{Barnes:1987, Warren:1992}, a dimensionless number that measures the degree of rotational 
support of the halo.  Following \cite{Bullock:2001}, we define it as
\begin{equation}
\lambda = \frac{L_{200}}{\sqrt{2}M_{200}V_{200}R_{200}}
\end{equation}
where the quantities the total angular momentum $L$, the total mass
$M$, the circular velocity $V$, and the radius $R$ are all taken as defined by \Eq{eq:virial_mass_definition},
with $\Delta=200$; in cosmological simulations, the distribution
of this parameter is found to be described as lognormal \citep[e.g.][]{Barnes:1987,
  Warren:1992, Cole:1996, Gardner:2001, Bullock:2001, Maccio:2007,
  Maccio:2008, Munoz-Cuartas:2011}
\begin{equation}
P(\lambda) = \frac{1}{\lambda\sigma_0^2\sqrt{2\pi}} \exp \left[ - \frac{\ln^2(\lambda/\lambda_0)}{2\sigma_0^2}\right] \ ,
\end{equation}
even though some authors \citep[e.g.][]{Bett:2007} claim that this should be slightly modified.

Due to the non-Gaussian nature of this distribution, instead of the
average value we plot in \Fig{img:spin} the median value of the spin parameter
${\lambda}$ as a function of halo mass.  A weak
negative correlation of spin to the halo mass can be observed here 
for haloes above $8\times10^{12}$\hMsun, (as
noted for instance by \cite{Maccio:2007} and \cite{KnebePower:2008})
while the relation is positive below that threshold.
However, \cde\ models have on average a
higher value (per mass bin) compared to \ude\ and \LCDM. Albeit small,
this increase in ${\lambda}$ is clearly a coupling related effect, the
magnitude of which is directly proportional to the value of $\beta$.
Given the small error bars (due to the large number of objects used in
this analysis) we are confident that this is a real effect.  Moreover, a
similar result has been found by \cite{Hellwing:2011}, that also
claimed to have observed a link between fifth force and larger $\lambda$s. 

A deeper investigation of the physical link between the coupling and
increased rotational support is left to an upcoming work (Carlesi et
al., in prep.) where the \textit{evolution} of different parameters
under different cosmologies will be analyzed.  For the moment it is
important to note that there appears to be some evidence of a link
between the coupling strength of the fifth force and the corresponding
degree of rotational support in dark matter haloes.
\begin{figure}
\includegraphics[width=8cm]{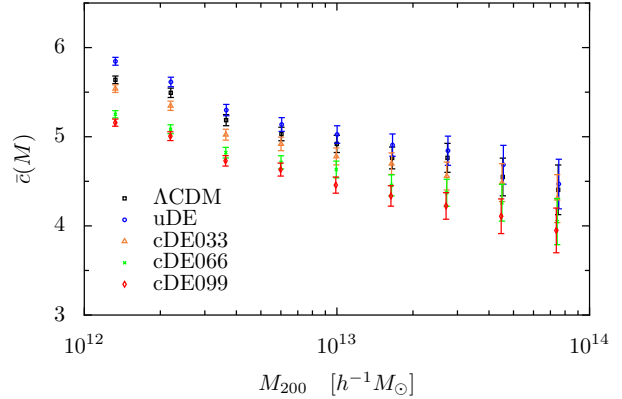}
\caption{\small Halo mass concentration relation at $z=0$,  where the median concentration
per mass bin is plotted.
Compared to \LCDM, \cde\ cosmologies show a systematically lower value of $c$ for all mass bins, 
whereas for \ude\ it is larger.}
\label{img:cm}
\end{figure}

\subsubsection{Concentration}

\begin{table*}
\caption{
Best fit values for the mass concentration relation 
for haloes with $M>10^{12}$\hMsun.}
\label{tab:c-M}
\begin{center}
\begin{tabular}{cccccc}
\hline
Model	 & \LCDM 	 & \ude 	  & \cde033 	    & \cde066 & \cde099 \\
\hline 
$c_0$	 &  $4.31 \pm0.06  $ & $ 4.48 \pm0.07 $ &  $4.18 \pm0.06 $  &  $3.98 \pm0.08$  & $ 3.86 \pm0.06 $\\
$\gamma$ &  $-0.088\pm0.007$ & $-0.093\pm0.009$ & $-0.083\pm0.009 $ & $-0.084\pm0.009$ & $-0.079\pm0.008 $\\
\hline
\end{tabular}
\end{center}
\end{table*}

Dark matter density profiles can be described by a Navarro-Frenk-White (NFW) profile \citep{NFW:1996},
of the form
\begin{equation}
\rho(r) = \frac{\rho_0}{\frac{r}{r_s}\left(1+\frac{r}{r_s}\right)^{2}}
\label{eq:nfw}
\end{equation}
where the $r_s$, the so called scale radius, and the density $\rho_0$ are in principle two 
free parameters that depend on the particular halo structure.  
Using \Eq{eq:nfw} we can define
\begin{equation}\label{eq:conc}
c = \frac{r_{200}}{r_s}
\end{equation}
which is the \emph{concentration} of the halo, relating the
radius $r_{200}$ to the scale radius $r_s$.
Fitting \Eq{eq:nfw} to our halo sample we observe that no 
substantial difference can be seen in the different simulations, that
is, the NFW formula describes (on average) equally well
dark matter halo profiles in \LCDM\ as in the other (coupled) dark
energy models. 
While this is in contrast with the early findings of \cite{Maccio:2004}, 
it is however in good agreement with the subsequent works of 
\citet{Baldi:2010a} and \cite{Li:2011}, who also found the NFW profile
to be a valid description of DM haloes in interacting cosmologies.
Thus, defining concentrations using \Eq{eq:conc} will not pose
any problems nor introduce any systematic effect due to the fact
that the NFW profile might only be valid for \LCDM\ dark matter
haloes.

In \Fig{img:cm} we now show the median concentration for objects in a
certain mass bin: \cde\ cosmologies have a smaller concentration than
\LCDM\, i.e. the larger the $\beta$ the smaller the $c$; whereas the
opposite is true for the \ude\ model.
This can partly be explained by the fact that concentrations are
related to the formation time of the halo, since structures that
collapsed earlier tend to have a more compact centre due to the fact
that it has more time to accrete matter from the outer parts.
Dynamical dark energy cosmologies generically imply larger $c$ values
as a consequence of earlier structure formation, as found in works
like those by \cite{Dolag:2004}, \cite{Bartelmann:2006} and
\cite{Grossi:2009}. In fact, since the presence of early dark energy
usually suppresses structure growth, in order to reproduce current
observations we need to trigger an earlier start of the formation
process, which on average yields a higher value for the halo
concentrations.  However, as explained in
\cite{Baldi:2010a}, smaller concentrations in \cde\ models are not related to the 
formation time of dark matter halos, but to the fact that one of the effects of 
coupled quintessence is to effectively act as a positive friction term.
This means that dark matter particles have an increased kinetic energy, which
moves the system out of virial equilibrium and causes a slight expansion, 
resulting in a lowering of the concentration.

In the hierarchical picture of structure formation,
concentrations are usually inversely correlated to the halo mass as
more massive objects form later; $N$-body simulations
\citep{Dolag:2004,Munoz-Cuartas:2011, Prada:2012} and observations
\citep{Comerford:2007,Okabe:2010,Sereno:2011} have in fact shown that
the relation between the two quantities can be written as a power law
of the form
\begin{equation}\label{eq:c-M}
c(M) = c_0 \left(\frac{M_{200}}{10^{14}M_{\odot}h^{-1}}\right)^{\gamma}
\end{equation}
where $\gamma$ and $c_0$ can have explicit parametrizations as
functions of redshift and cosmology \citep[see]
[]{Neto:2007,Munoz-Cuartas:2011, Prada:2012}. When we fit our
halo sample to this relation using $c_0$ and $\gamma$ as free
parameters we obtain the best-fit values as shown in
\Tab{tab:c-M}. Our values are qualitatively in good agreement with the
ones found by, for instance, \cite{Maccio:2008} and
\cite{Munoz-Cuartas:2011} for \LCDM; but we do find some tension with
the findings of \cite{Prada:2012}. However, since they use a different
algorithm for the determination of $c$ (which, according to them,
leads to higher concentration values) and a different $\sigma_8$
normalizations we cannot directly compare our results to theirs.  On
the other hand, \ude\ values are generally in agreement with
\cite{Dolag:2004}, \cite{DeBoni:2013} although in both cases there are
again some discrepancies in the $c_0$ best-fit result, most probably
due to the much different $\sigma_8$ used in their simulations. For
\cde\ we cannot directly compare our concentration-mass relation to
the one obtained by \cite{Baldi:2010a} since they do not provide any
fit to \Eq{eq:c-M}. 
\begin{figure*}
\begin{center}
\begin{tabular}{c}
\subfigure[\LCDM]{\includegraphics[width=12cm]{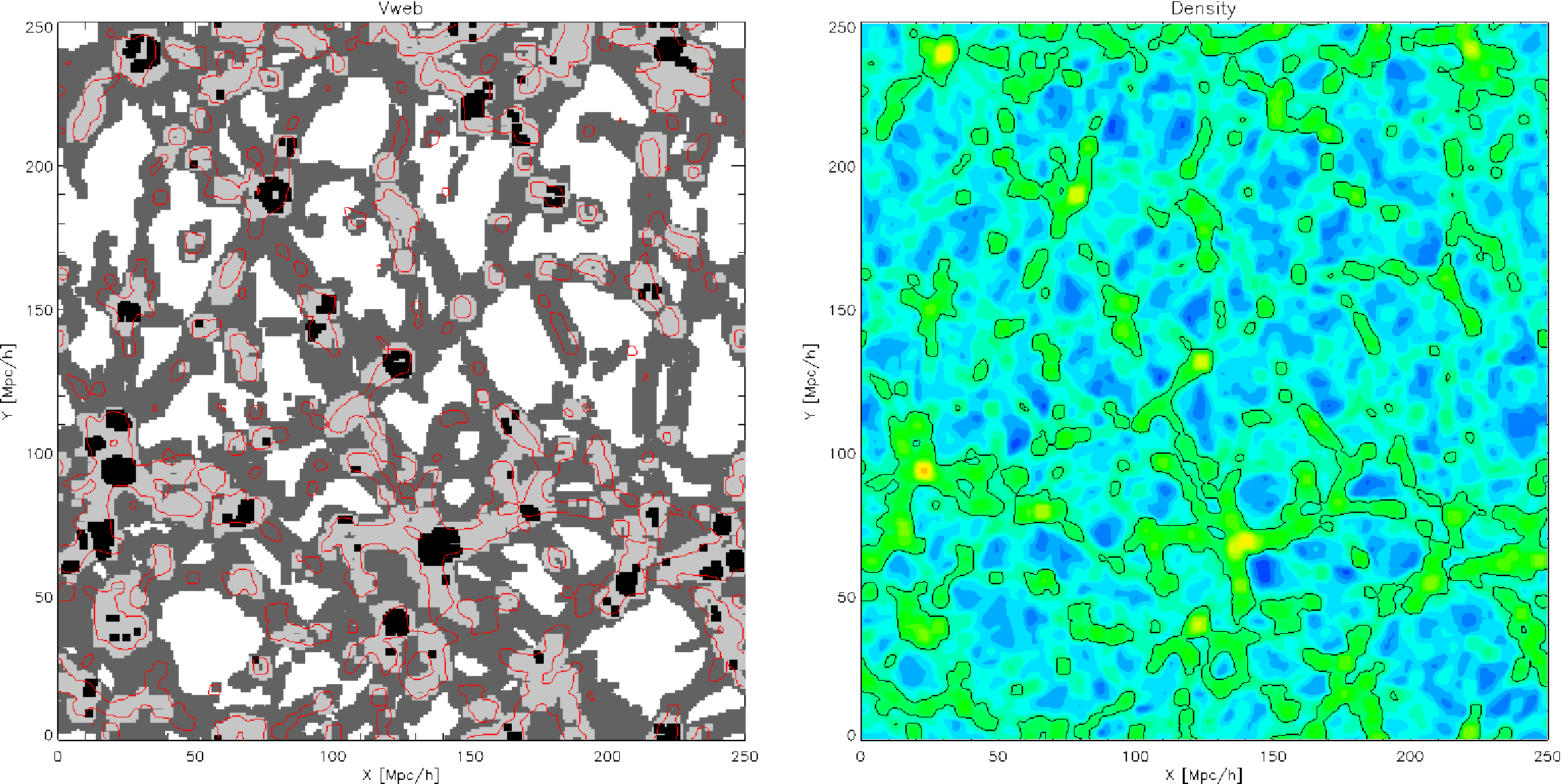}}\\
\subfigure[\ude]{\includegraphics[width=12cm]{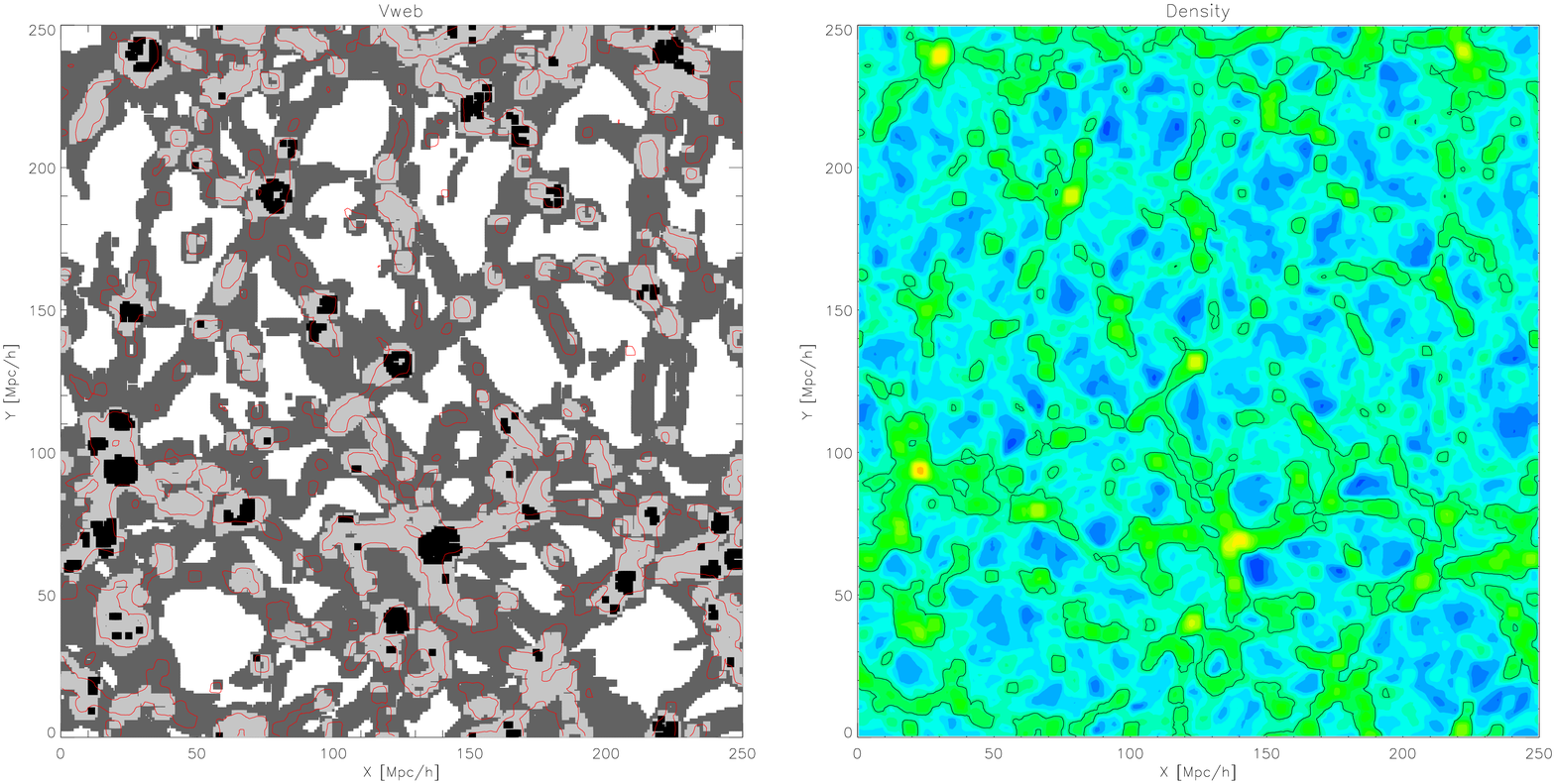}}\\
\end{tabular}
\caption{\small Cosmic web and matter density field plots 
for \LCDM\ (upper panel) and \ude (lower panel).
In the left panels of each pair we plot voids (white points), sheets (dark grey), filaments (light grey) and 
knots (black), while on top of them we depict red contours enclosing the regions where $\delta>1$.
The right panel of each pair show the colour coded logarithmic matter density, the black solid
contours again encompass overdense regions.
We notice that there's a very good overlap of overdense regions with filaments and knots, while
underdense ones can be identified with voids and sheets. 
}
\label{img:web}
\end{center}
\end{figure*}

\begin{figure*}
\begin{center}
\begin{tabular}{cc}
\subfigure[\cde033]{\includegraphics[width=12cm]{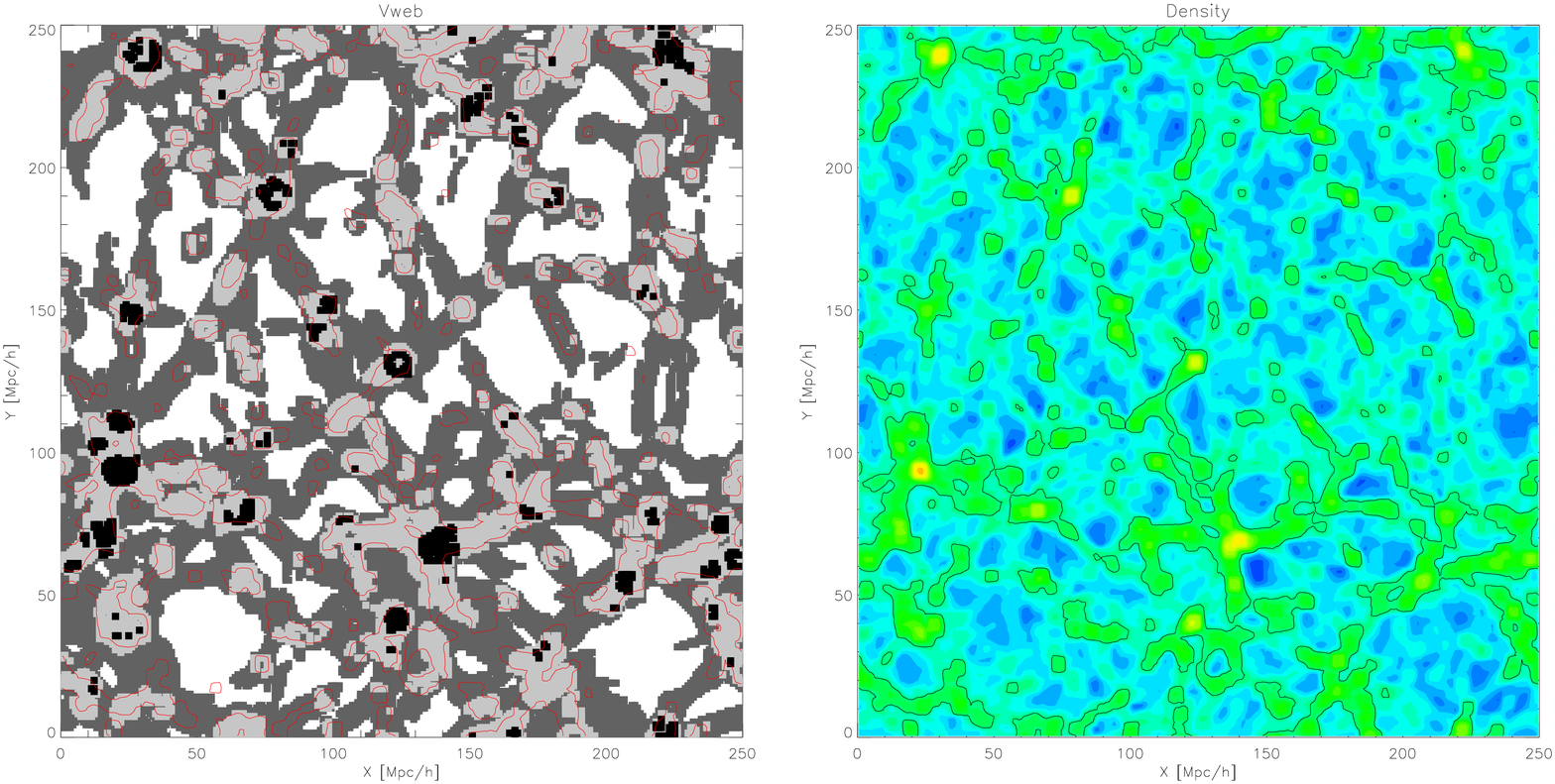}}\\
\subfigure[\cde066]{\includegraphics[width=12cm]{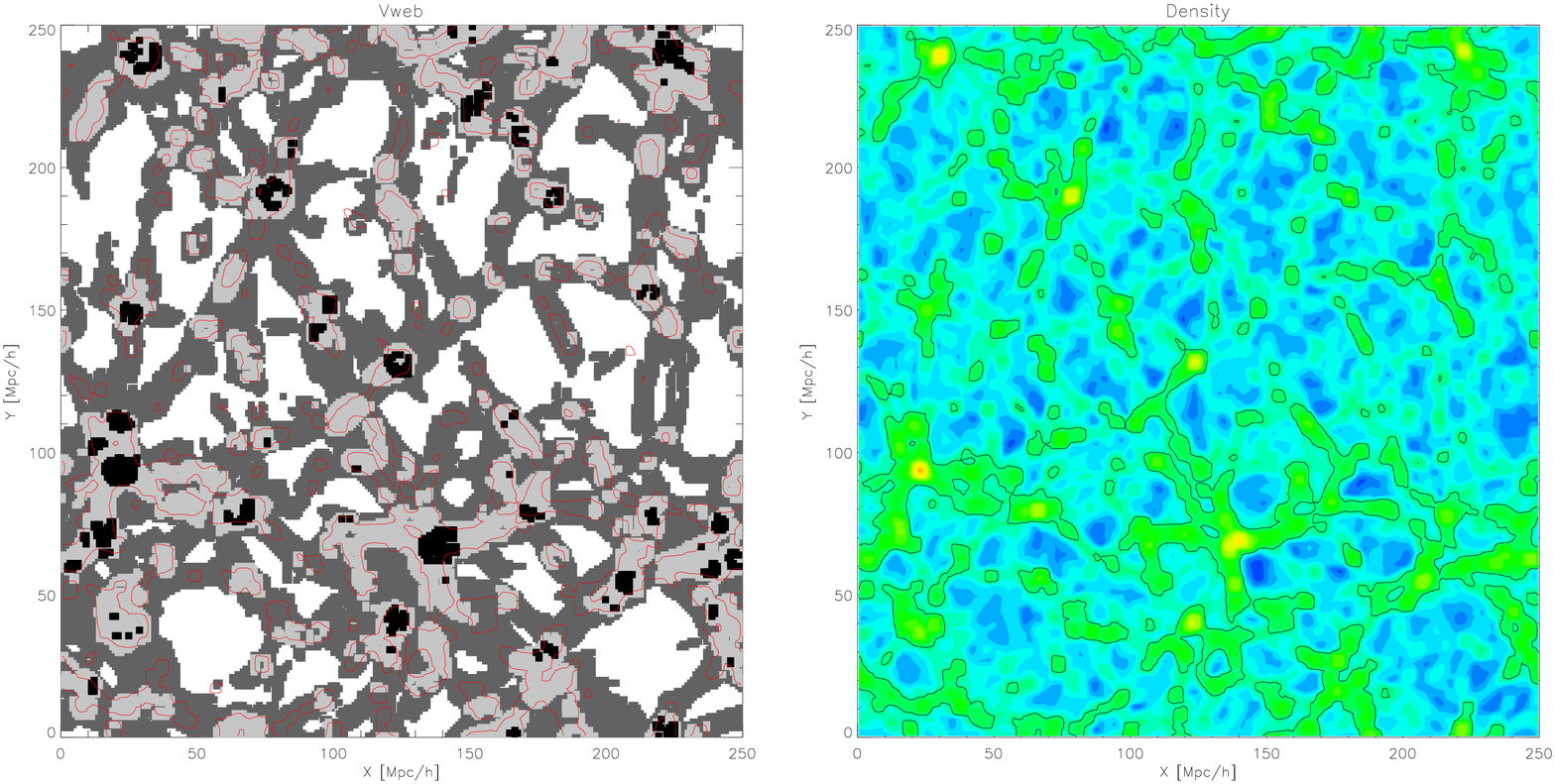}}\\
\subfigure[\cde099]{\includegraphics[width=12cm]{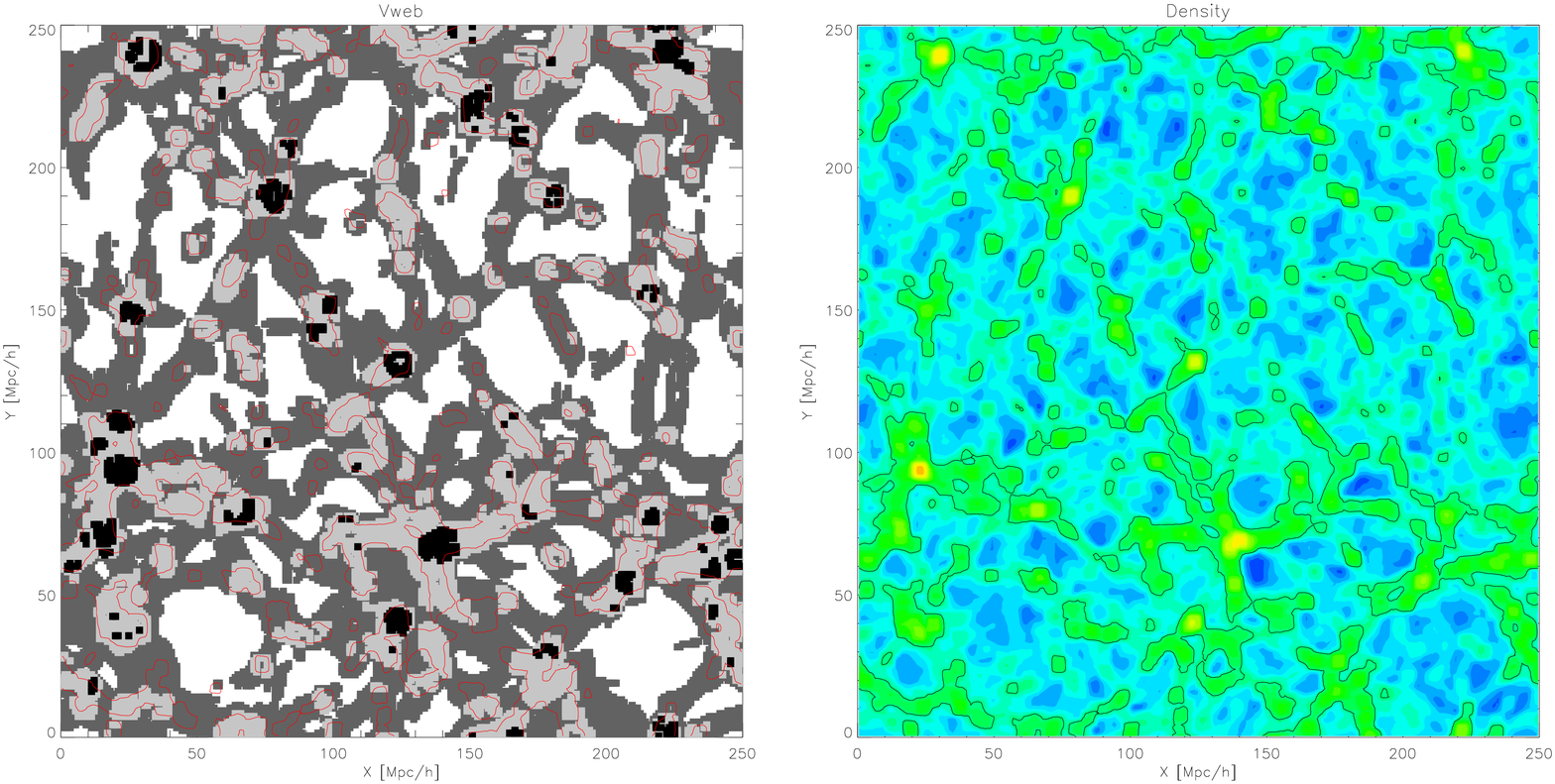}}
\end{tabular}
\caption{\small
\small 
Same as \Fig{img:web} for \cde033 (upper panel), \cde066 (middle panel) and \cde099 (lower panel).}
\label{img:web_cont}
\end{center}
\end{figure*}

\begin{figure*} 
\begin{center}
\begin{tabular}{ccc}
\includegraphics[width=5.5cm]{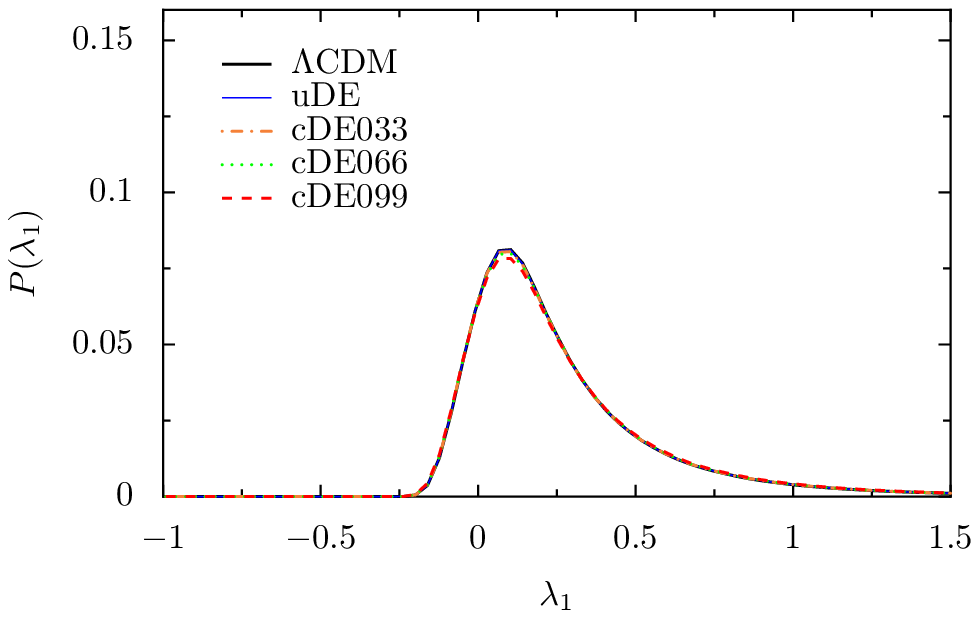} &
\includegraphics[width=5.5cm]{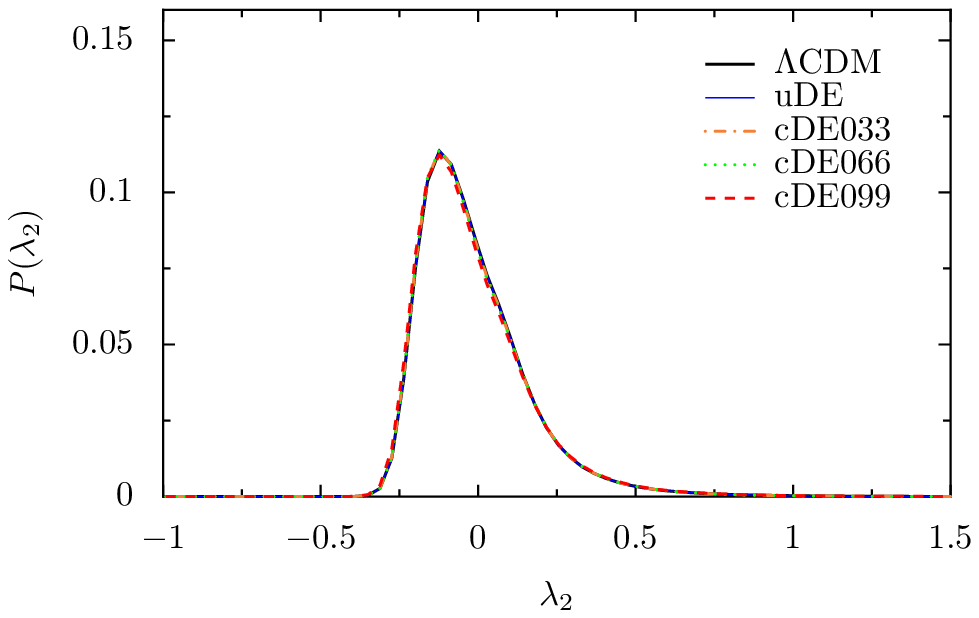} &
\includegraphics[width=5.5cm]{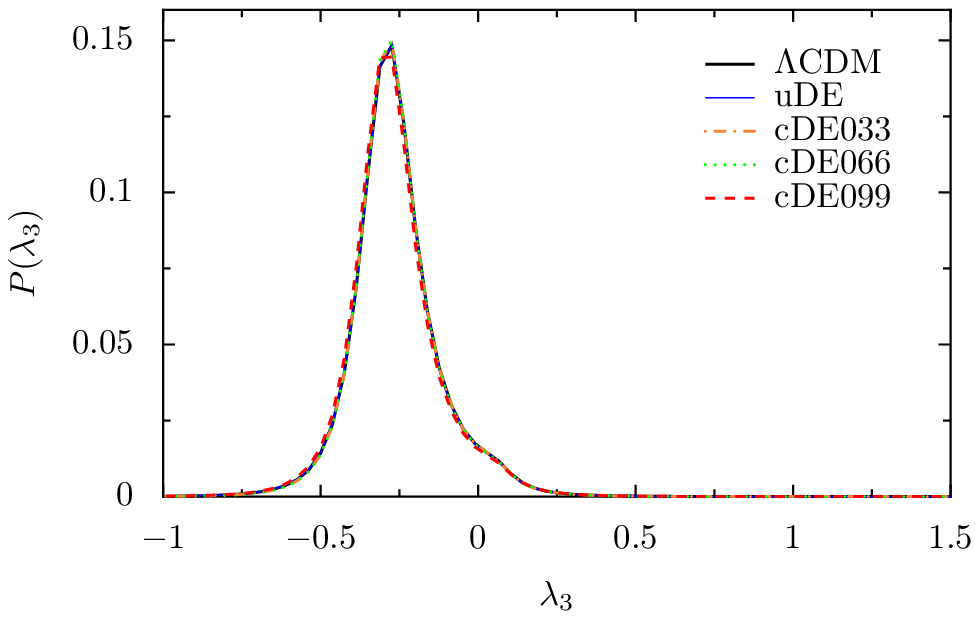} \\
\end{tabular}
\caption{Probability distributions for the eigenvalues of the velocity shear tensor at all nodes. 
At every node we assume $\lambda_1 > \lambda_2 > \lambda_3$. The distributions are almost identical
for all simulations and eigenvalues, except for a progressively lower peak of $P(\lambda_1$ (left panel) 
for coupled models.
}
\label{img:eigenvalues}
\end{center}
\end{figure*}

\section{Properties of the cosmic web}\label{sec:web_res}

\begin{table}
\caption{Fraction of total dark matter mass for different node type in each simulation.}
\label{tab:massfrac}
\begin{center}
\begin{tabular}{lccccc}
\hline
cell type	 & \LCDM & \cde & \cde033 & \cde066 & \cde099 \\
\hline 
void	 & 0.103 & 0.103 & 0.102 & 0.103 & 0.102 \\
sheet	 & 0.343 & 0.343 & 0.344 & 0.344 & 0.343 \\
filament & 0.437 & 0.438 & 0.443 & 0.442 & 0.443 \\
knot 	 & 0.116 & 0.115 & 0.109 & 0.111 & 0.118 \\
\hline
\end{tabular}
\end{center}

\caption{Fraction of total gas mass for different node type in each simulation.}
\label{tab:gasfrac}
\begin{center}
\begin{tabular}{lccccc}
\hline
cell type	 & \LCDM & \cde & \cde033 & \cde066 & \cde099 \\
\hline 
void	 & 0.103 & 0.103 & 0.103 & 0.104 & 0.102 \\
sheet	 & 0.349 & 0.348 & 0.348 & 0.347 & 0.346 \\
filament & 0.449 & 0.450 & 0.450 & 0.449 & 0.453 \\
knot 	 & 0.097 & 0.097 & 0.097 & 0.098 & 0.098 \\
\hline
\end{tabular}
\end{center}

\caption{Volume filling fractions of different cell types for all the simulation set.}
\label{tab:volumefrac}
\begin{center}
\begin{tabular}{lccccc}
\hline
cell type 	 & \LCDM & \cde & \cde033 & \cde066 & \cde099 \\
\hline 
void   	 & 0.337 & 0.338 & 0.338 & 0.337 & 0.334 \\
sheet	 & 0.460 & 0.456 & 0.460 & 0.460 & 0.461 \\
filament & 0.185 & 0.184 & 0.185 & 0.185 & 0.186 \\
knot 	 & 0.017 & 0.017 & 0.017 & 0.017 & 0.018 \\
\hline
\end{tabular}
\end{center}
\end{table}

We now turn to the study of the cosmic web, as defined in \Sec{sec:web},
in \LCDM, \ude, \cde033, \cde066 and \cde099. 
In \Fig{img:web} and
\Fig{img:web_cont} we give a visual impression of the web
classification (left) and the underlying dark matter density field
(right) for a slice of thickness one cell (i.e. $0.97$\hMpc) using a
logarithmic colouring scheme for the density.
From Figs.~\ref{img:web}~\&~\ref{img:web_cont} it is evident that
there is, in general, a very close correspondence between $\delta>1$
and filamentary and knot-like regions; just like between $\delta<1$
and void and sheet-like ones, so that the kinetic classification does
provide in general a faithful description of the underlying density
distribution -- as shown in \citet{Hoffman:2012}. 
Nonetheless, a minor number of cells do indeed violate this
principle.  In fact, as also noted by \cite{Hoffman:2012}, in a very
limited number of cases it happens that, for cells placed in the
interior of a of a large dark matter halo, the velocity field will be
determined by the motion of its virialized particles and not
reflecting the cosmic web, respectively. On top of that, we must not
forget that the freedom in the choice of the threshold $\lambda_{th}$,
and the fixed spacing of the grid account for the fact that on scales
smaller than $0.97$\hMpc\ we cannot properly resolve the complex shape
of the web, which would probably require a more flexible grid
implementation \citep{Platen:2011}.  However, all these shortcomings
do not seriously invalidate this description, as the number of such cells
is generally small (for example, points defined as voids with
$\delta>1$ sum up to less than $1\%$ of the total in all simulation,
and independent of the simulation).  In fact, the latter is the most
important condition that we need to ensure, so that the existence of
small biases disappears when considering ratios to \LCDM, which is at
the core of the analysis we are carrying.

In Tabs.~\ref{tab:massfrac},~\ref{tab:gasfrac}~\&~\ref{tab:volumefrac} we show the mass and
volume filling fractions as a function of cell type and
cosmologies. These values are estimated simply summing all masses and
volumes contained in cells belonging to the same kind of environment.
What is clear by looking at these results is that the general
structure of the cosmic web is almost left unchanged across models. In
fact, discrepancies among different cosmologies are much less than
$1\%$ in this regard, thus making it hard to detect deviations from
\LCDM\ by simply considering the volume and the mass associated to the
various kinds of environment.  The same conclusion can be drawn if we
look at \Fig{img:eigenvalues}, which shows the distributions of the
three $\lambda_{1,2,3}$ eigenvalues of $\Sigma_{\alpha\beta}$, that
appear to be identical and thus provide no leverage to distinguish
the models under investigation here.

The gas distribution through the different node types seems also to be largely unaffected
by the different cosmology: As we can see from \Tab{tab:gasfrac}, the mass fractions of gas
are substantially identical throughout all the models, without any significant discrepancy.
Comparing to the distribution of dark matter, we do notice a slight increase in the 
fraction of gas belonging to sheets and filaments paralleled by its reduction on knots,
a pattern which is observable in all the models to the same extent.

We remark that our results for
\ude\ agree with \cite{Bos:2012}, who also found that 
quintessence cosmologies with Ratra-Peebles potentials 
do not lead to significant changes in the general
properties of the cosmic web.
We also emphasize  that 
our findings relative to void regions are largely independent of the choice of $\lambda$.
Using different threshold values we have been able to test this and see
that void distributions are affected to the same degree in all the different models, confirming
this particular result does not depend on our $\lambda_{th}$.

\begin{figure*}
\begin{center}
\begin{tabular}{cc}
\includegraphics[height=6cm]{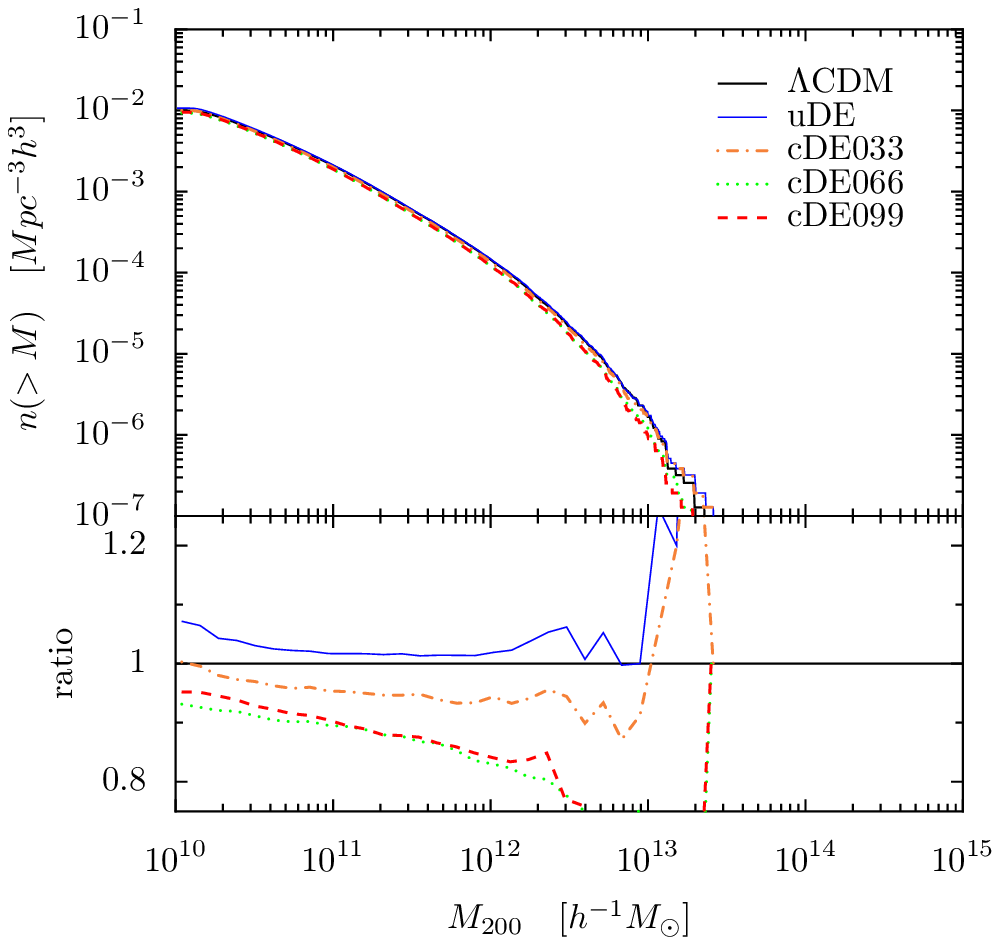} &
\includegraphics[height=6cm]{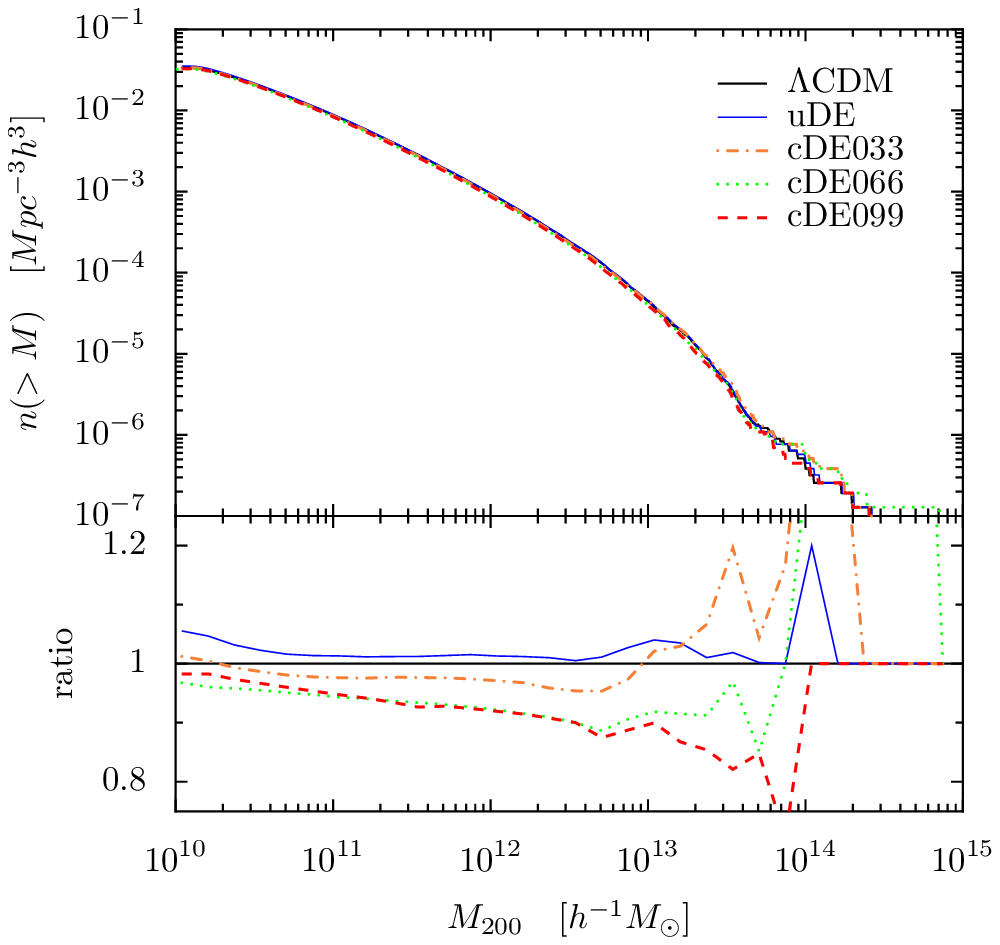} \\
\includegraphics[height=6cm]{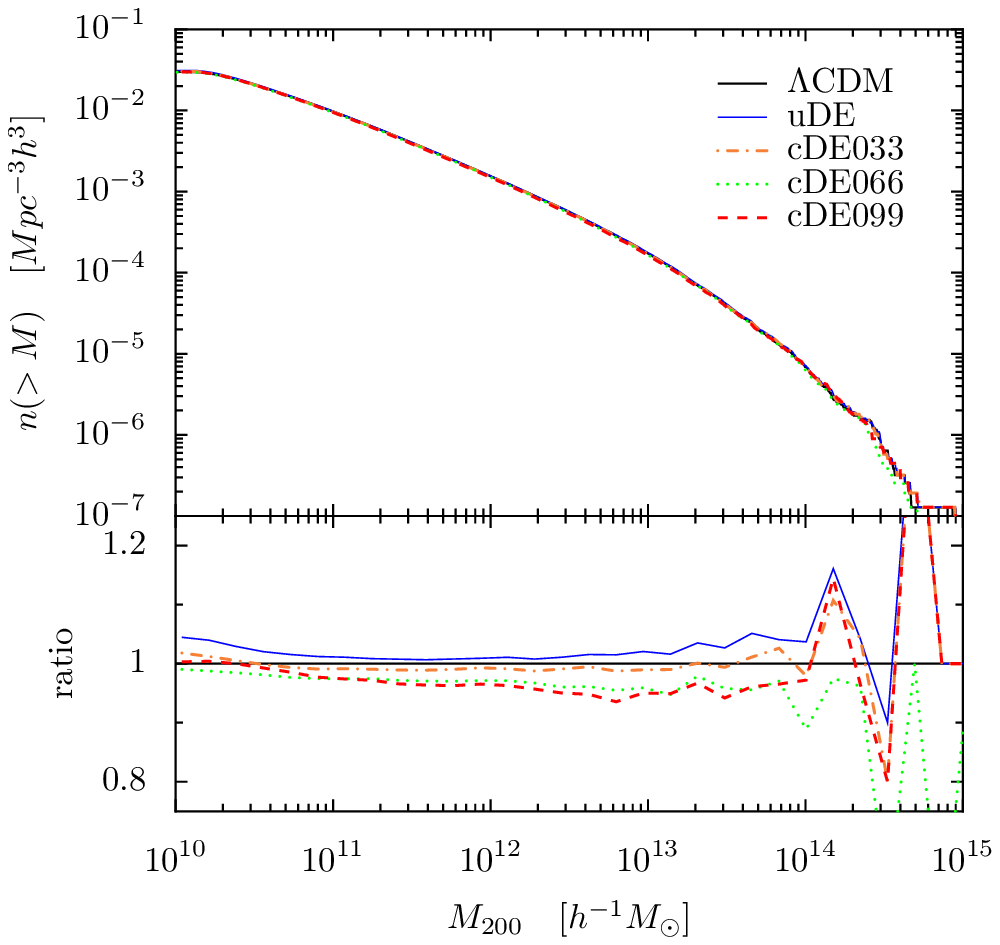} &
\includegraphics[height=6cm]{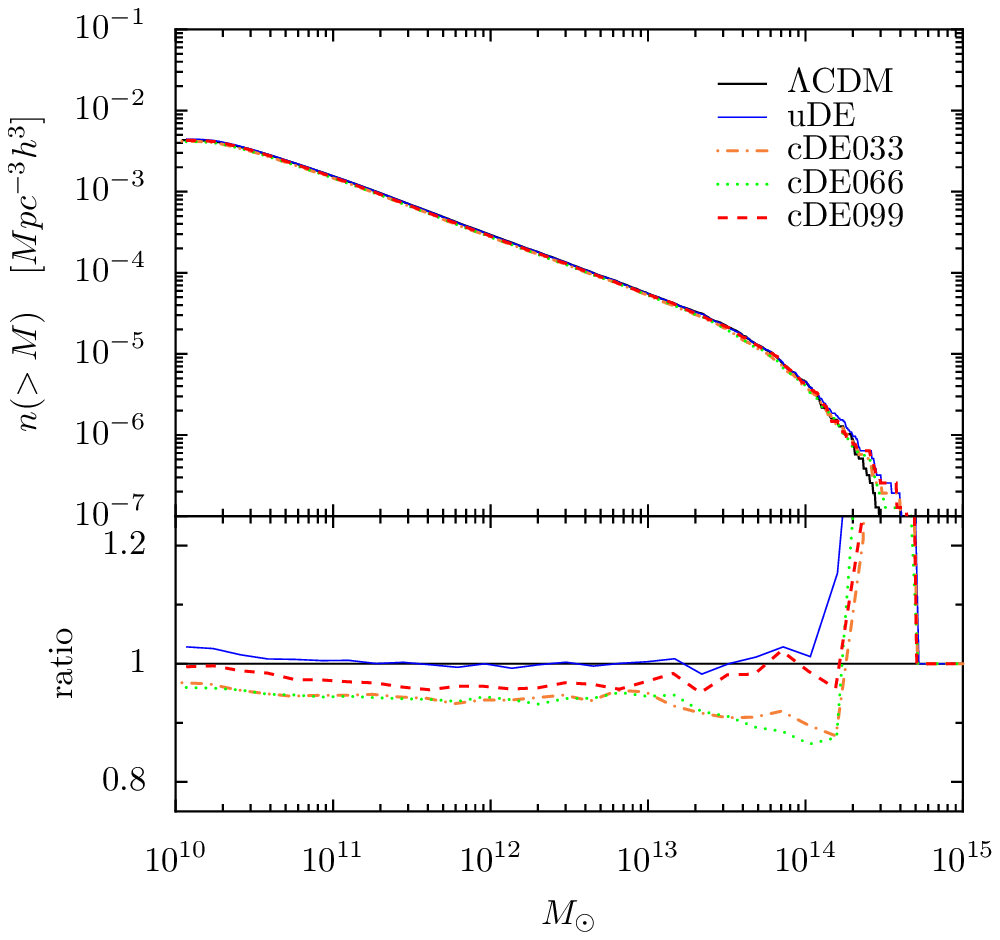} \\
\end{tabular}
\caption{
Upper panels: halo mass function in voids (left) and sheets (right). 
Lower panels: halo mass function in filaments (left) and knots (right). }
\label{img:mass_env}
\end{center}
\end{figure*}

\begin{figure*}
\begin{center}
\begin{tabular}{cc}
\includegraphics[height=6cm]{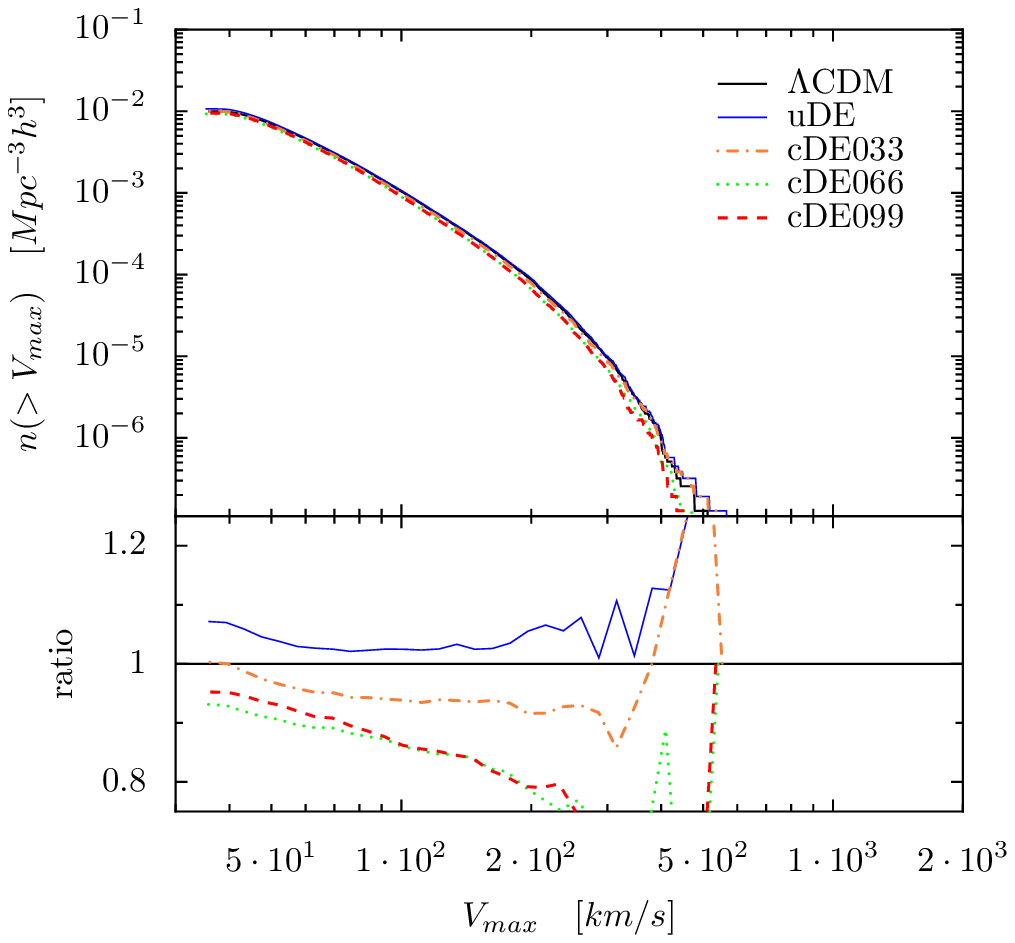} &
\includegraphics[height=6cm]{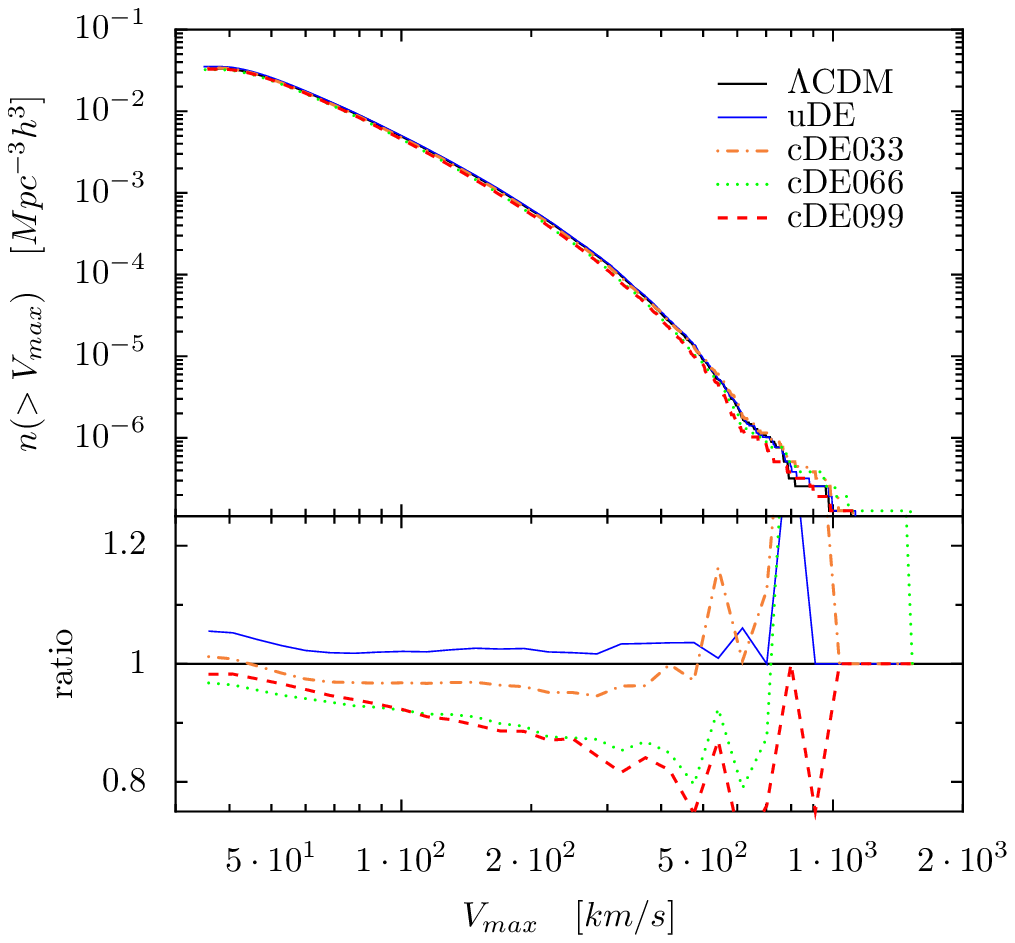} \\
\includegraphics[height=6cm]{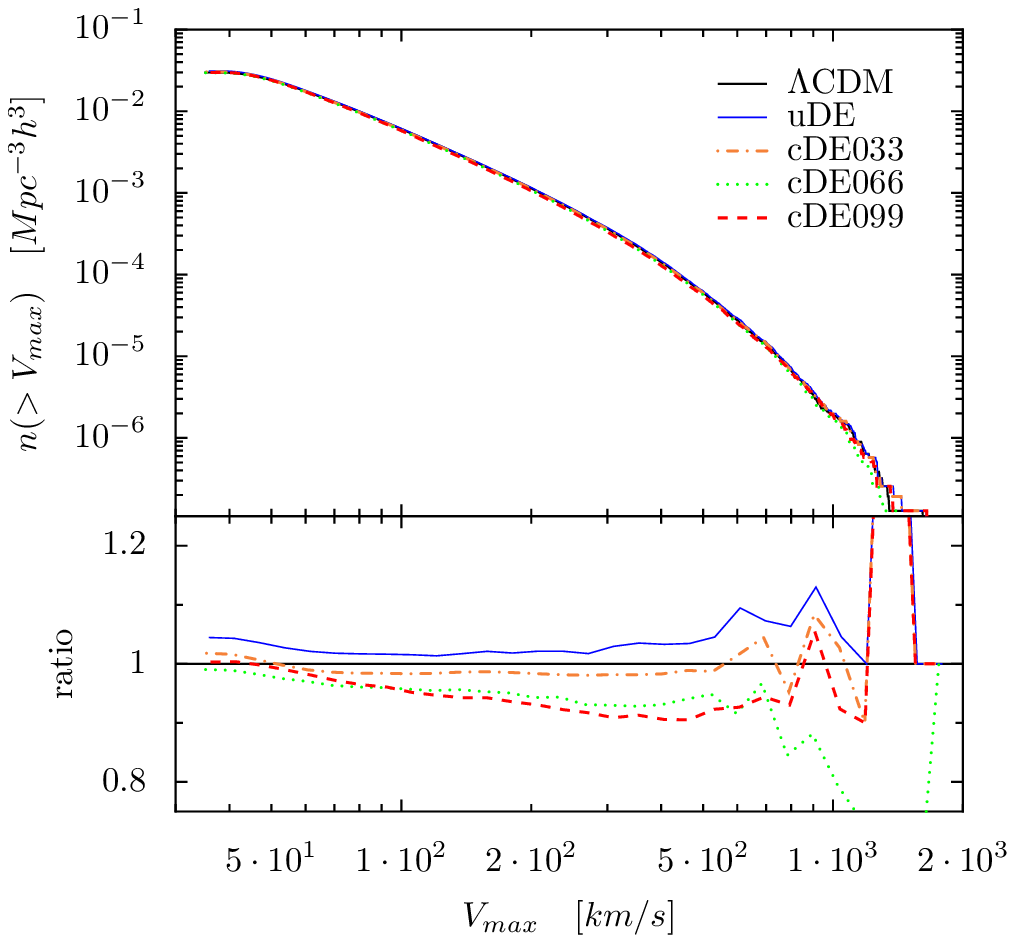} &
\includegraphics[height=6cm]{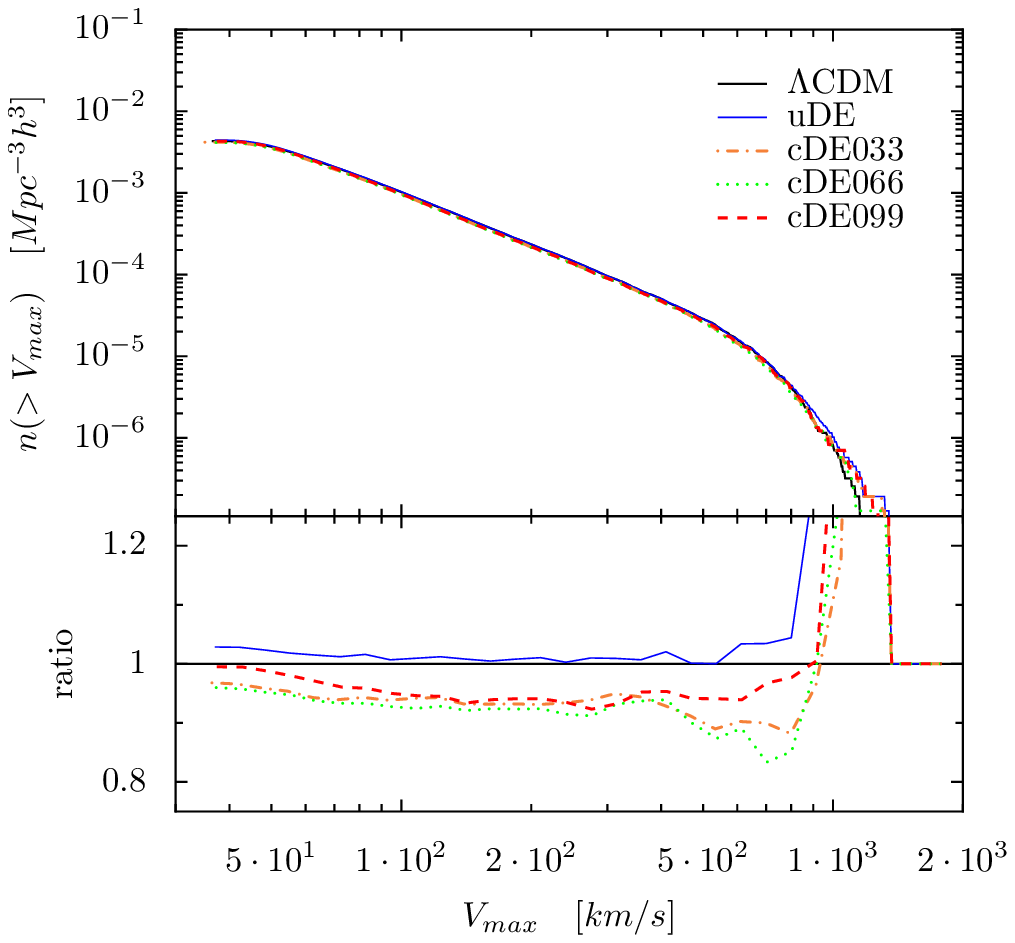} \\
\end{tabular}
\caption{
Upper panels: halo velocity function in voids (left) and sheets (right).
Lower panels: halo velocity function in filaments (left) and knots (right).}
\label{img:vel_env}
\end{center}
\end{figure*}

\begin{figure*}
\begin{center}
\begin{tabular}{cc}
\includegraphics[width=8cm]{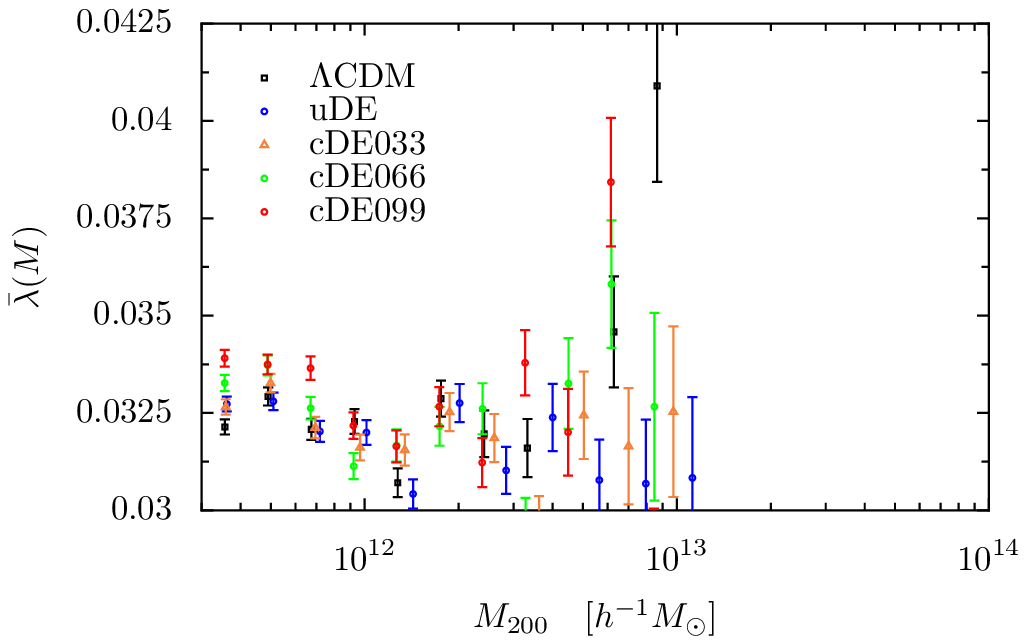} &
\includegraphics[width=8cm]{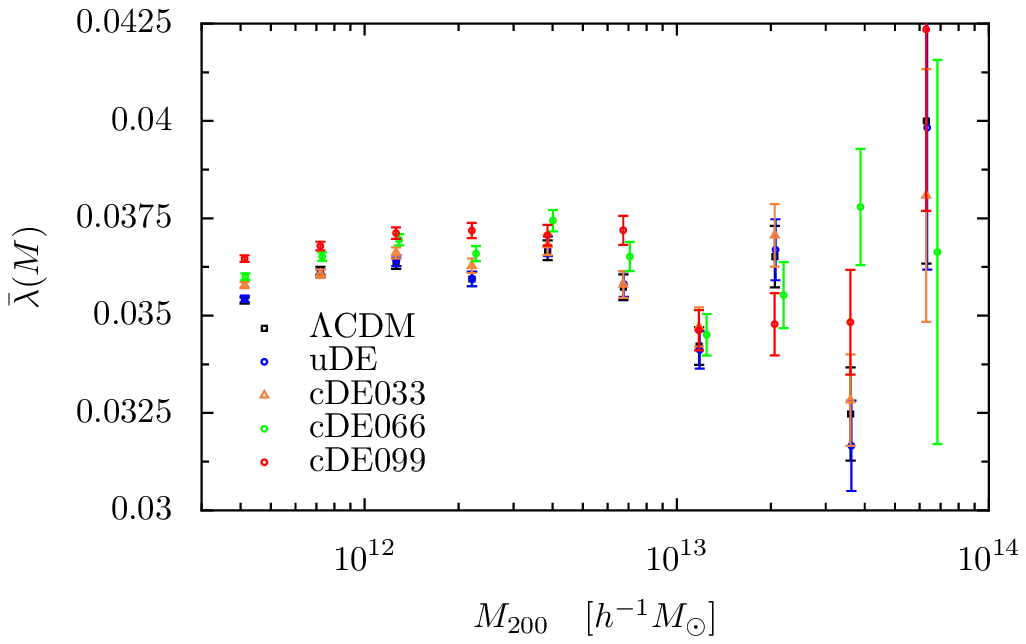} \\
\includegraphics[width=8cm]{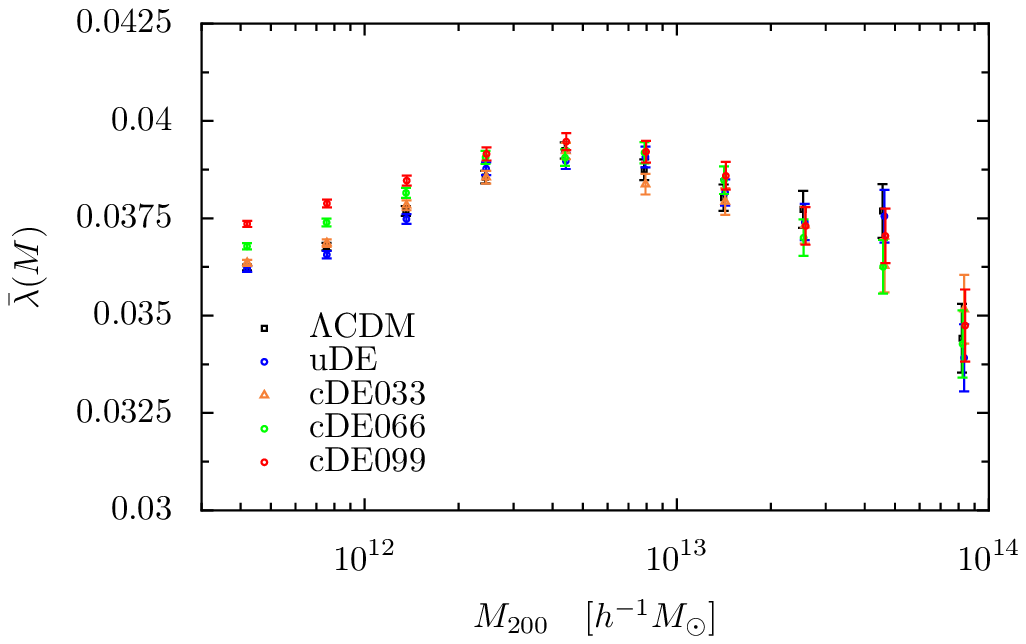} &
\includegraphics[width=8cm]{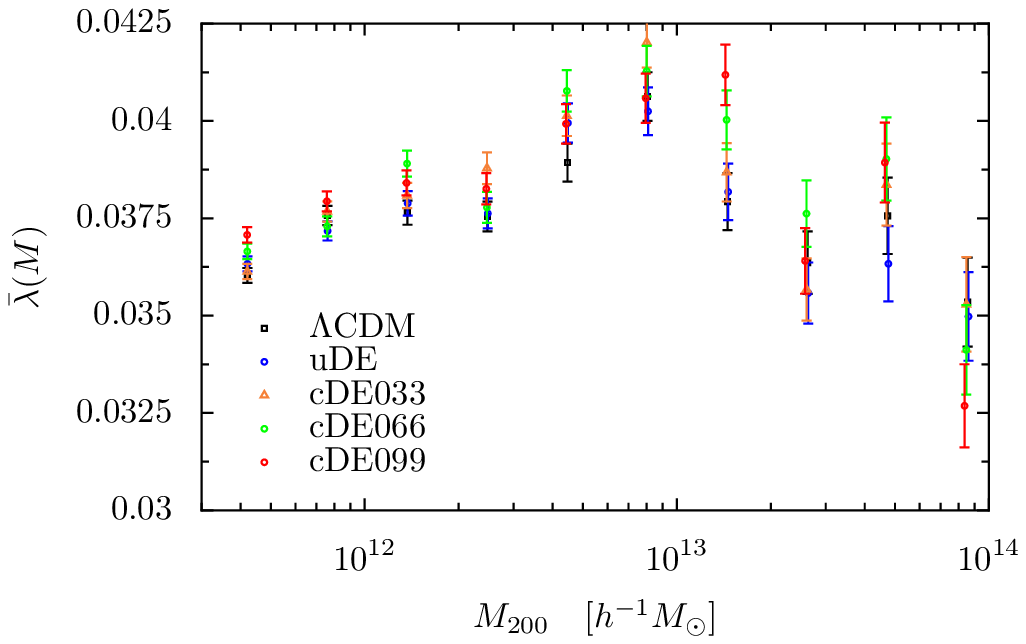} \\
\end{tabular}
\caption{\small Median of the spin parameter for haloes located in voids (upper left panel), 
sheets (upper right panel) filaments (lower left panel) and knots (lower right panel).}
\label{img:spin_env}
\end{center}
\end{figure*}

\begin{figure*}
\begin{center}
\begin{tabular}{cc}
\includegraphics[width=8cm]{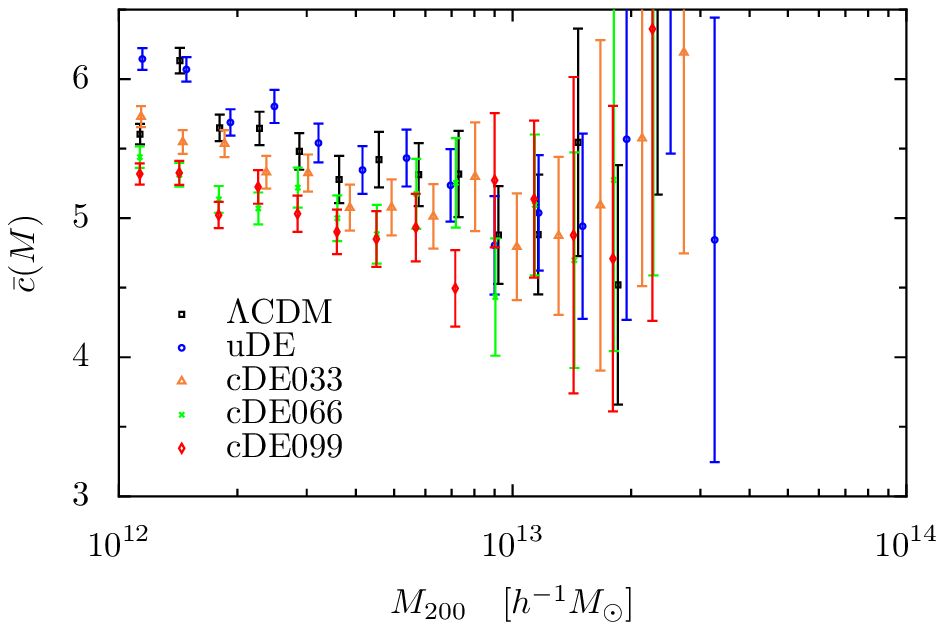} &
\includegraphics[width=8cm]{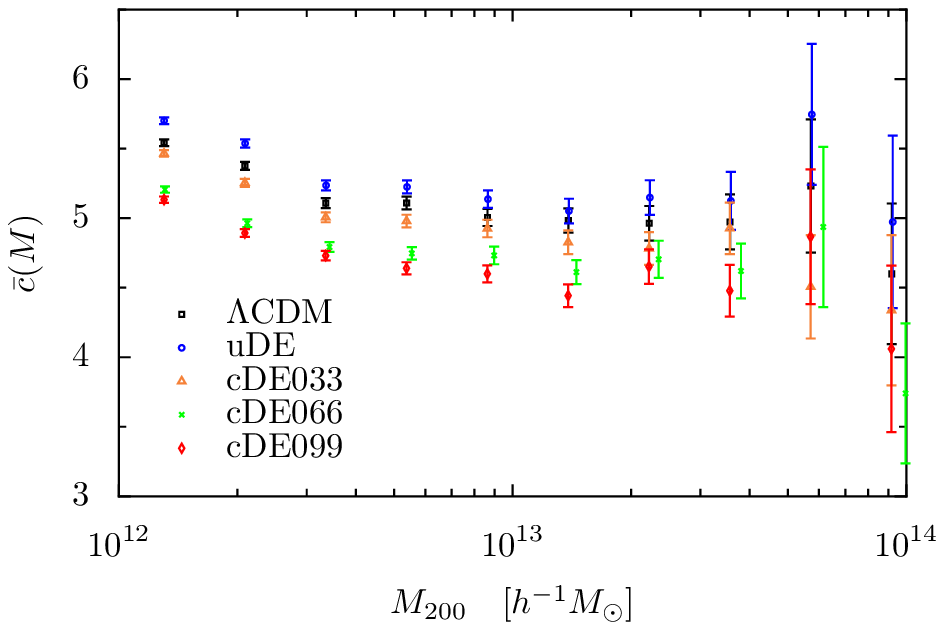} \\
\includegraphics[width=8cm]{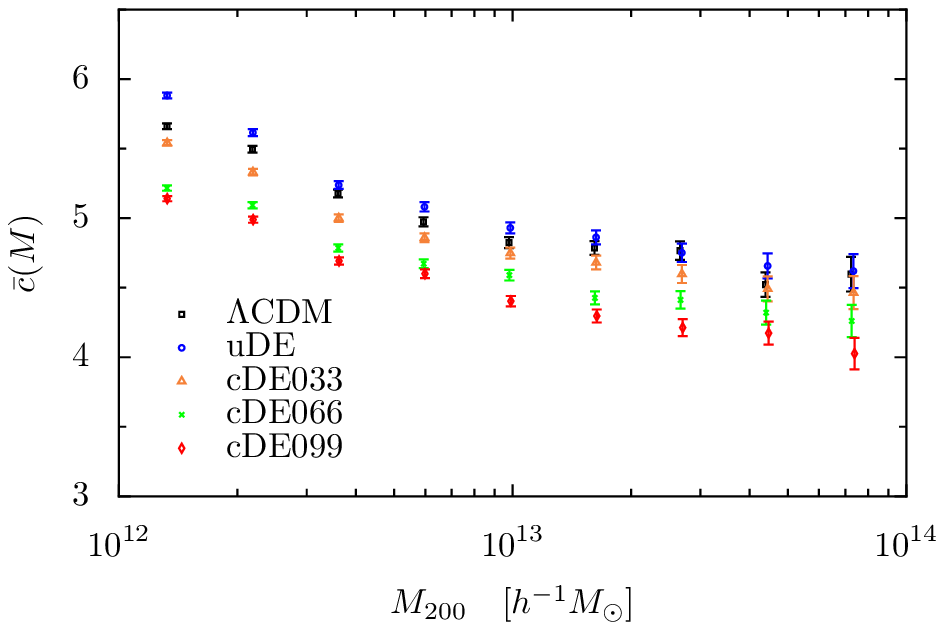} &
\includegraphics[width=8cm]{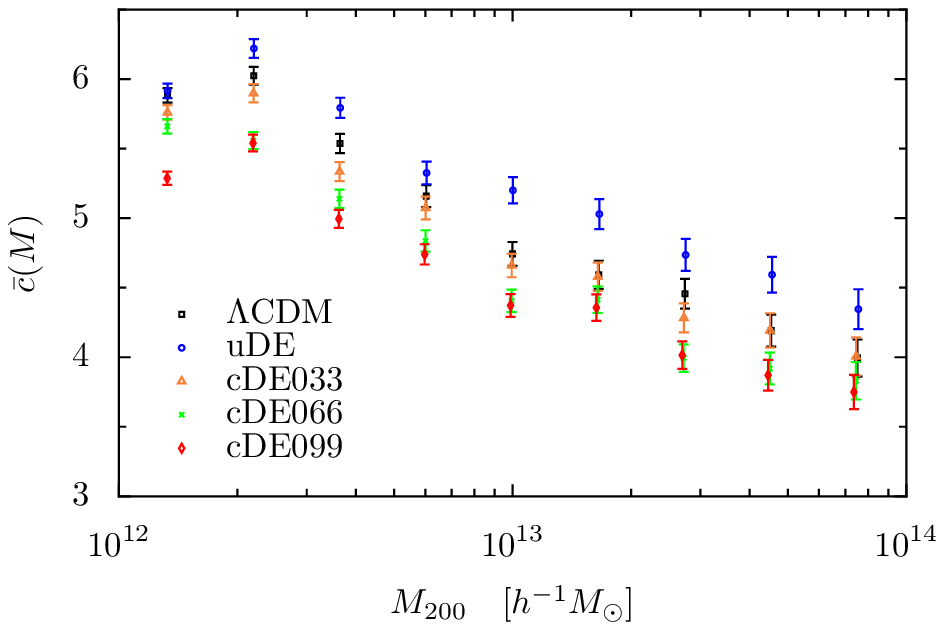} \\
\end{tabular}
\caption{\small Average concentration for haloes located in voids (upper left panel), sheets (upper right panel), 
filaments (lower left panel) and knots (lower right panel).}
\label{img:cm_env}
\end{center}
\end{figure*}

\section{Halo properties in different environments}\label{sec:haloweb_res}
We now turn to the study of halo properties classified according to
their environment; this kind of analysis has already been done for
\LCDM\ using both geometrical \citep[e.g.][]{Avila-Reese:2005,
  Maccio:2007} and dynamical \citep[e.g.][]{Hahn:2007, Libeskind:2012,
  Libeskind:2013} web classifications, finding in general a
correlation between halo properties such as spin and shape to its
surrounding environment.

Using the information from the halo catalogues we proceed to assign
each halo to the nearest grid point and build up four different halo
samples, one for each cell type.  Then we repeat the analysis
presented in \Sec{sec:halo_res} for the halo counts (velocity and
mass), spin and concentrations.  We will see that this kind of
separation of haloes enhances some of the differences already seen in
general among different cosmological models and is therefore of great
importance when trying to constrain more effectively coupled and
uncoupled scalar field cosmologies.

Voids and sheets are readily identified with underdense regions, as
has also been confirmed by the analysis presented in the previous
\Sec{sec:web_res}. And the fact that for these cells at most one
eigenvalue of the shear tensor has a value above $\lambda_{th}$ means
that in two or more spatial directions there is a net outflow of
matter, which is in turn associated with a matter density below the
average. For overdense regions (i.e. filaments and knots) there is a
net inflow of matter towards the center of the cell from at least two
directions. Following this we partition the subsequent study into
underdense regions on the one hand (using voids and sheets) and
overdense regions (i.e. filaments and knots) on the other.

Underdense regions in \LCDM\ are usually associated with lower spins
and slightly larger halo concentrations \citep{Maccio:2007},
raising the question whether this still holds for (coupled) dark
energy cosmologies. 

\begin{table*}
\caption{
Fraction of haloes above $10^{10}$\hMsun\ per environment type, in each cosmological model. 
}
\label{tab:n_halo_env}
\begin{center}
\begin{tabular}{lccccc}
\hline
Environment& \LCDM 	 & \ude 	  & \cde033 	    & \cde066 & \cde099 \\
\hline 
void &  $0.128$ & $0.131$ &  $0.127$  &  $0.123$  & $0.124$\\
sheet &  $0.431$ & $0.432$ &  $0.431$  &  $0.429$  & $0.428$\\
filament &  $0.384$ & $0.382$ &  $0.387$  &  $0.392$  & $0.391$\\
knot &  $0.055$ & $0.054$ &  $0.053$  &  $0.055$  & $0.056$\\
\hline
\end{tabular}
\end{center}
\end{table*}

\subsection{Halo number counts}
Even though, by definition, underdense regions are
less populated, non-negligible fractions of the total 
halo count can be still found in voids and sheets, as shown in \Tab{tab:n_halo_env}, 
ensuring that the samples used are reasonably
large, and allow us to draw credible conclusions.

\subsubsection{Underdense regions: voids and sheets}
We notice that in underdense regions (i.e. voids and sheets, shown in the upper two plots)
the trend persists that the number
of objects is smaller for \cde\ models than for \LCDM, something also
observed for the general halo sample.  However, it is important to
remark that singling out and counting the objects belonging to the
underdense parts only, we end up observing larger differences among
the models. This effect also appears to be much stronger in \cde\ than
in \ude. In fact, whereas the differences in number counts of objects
does not exceed $7\%$, when restricting halo counts to void regions
only, we can see that \cde\ models' underprediction is much larger and
peaks at $20\%$ (ignoring the higher mass ends, where only a small
number of objects is found).  It is also clear from \Fig{img:vel_env}
that while the sign of the effect is very similar in both voids and
sheets, its strength is slightly reduced in the latter type of web,
suggesting that there exists at least a mild dependence of this
phenomenon on the specific kind of environment.  Although we are not
showing it here, we have also carefully checked that this result is
substantially independent from the kind of $\lambda_{th}$ chosen.  In
fact, repeating our computation using higher threshold values, we see
that the magnitude of the effect does not change substantially.
The physical mechanism behind this effect is
understood and provides a consistent framework for interpreting our results.
In fact, as first explained by \citet{Keselman:2010} and subsequently confirmed by
\citet{Li:2011}, fifth forces enhance the gravitational pull towards the overdense regions, 
quickly evacuating matter from underdense regions. 
This causes these environments to have less structures, so that in the end 
the number of haloes left in voids will be comparatively smaller 
than in the non-interacting cases, as found in our simulations.

However, we need to make an additional remark on this result before proceeding to the 
next section. In fact, we note that our choice of the $\sigma_8$
normalization, which is taken to be the same at $z=0$, plays an important role in
the result just described. It is in fact known \citep[see e.g.][]{Baldi:2010}
that using a different normalization prescription (for instance, at
the redshift of the CMB for the matter density fluctuations), coupled
models end up predicting (in total) more objects than \LCDM.
Hence, at this stage we cannot completely disentangle the influence of our choice
of the normalization of the initial conditions from the genuine influence of the additional interaction.

\subsubsection{Overdense regions: filaments and knots}
Like in the case of underdensities, we notice for overdense regions (shown in the lower 
two plots) that the trend of
suppression which characterizes the general halo counting still holds,
even though now the strength of this effect is slightly smaller across
all cosmologies. This is not unexpected, since the effect seen in the
HMF discussed in \Sec{sec:halo_res} has to be obtained from a
combination of both underdense and overdense structures, and should
therefore result in an intermediate value for \cde\ halo
underproduction.  Again, we have checked that the chosen threshold for
the eigenvalues of the velocity shear tensor does not substantially
affect this conclusion.

We can therefore state that there is a progression towards smoothing out
the differences among different cosmologies while moving to
increasingly higher density regions. \emph{This is a very important result
that indicates that underdense regions should be the target of choice
when searching for the effects of additional long range
gravitational-like forces}.

This result is in line with what has been already found for other fifth-force cosmologies 
\citep[see][]{Martino:2009,Keselman:2010,Li:2012,Winther:2012}, where the 
environmental dependence and in particular the properties of voids were stressed
as powerful tests for additional interactions and modifications of standard Newtonian
gravity.  It is in fact well known that void properties are extremely sensitive to 
cosmology \citep{Lee:2007,Lavaux:2009,Bos:2012,Sutter:2012} and hence provide a powerful probe of 
alternative models.
In particular, when the extra coupling in the dark sector is weak (as in the cases
analyzed here) the complex evolution and phenomena that characterize the overdense regions
may conceal its imprints, while void regions, whose dynamics is comparatively simpler,
are expected to be more directly linked to the underlying cosmology.

\begin{table*}
\caption{Best fit values for the concentration-mass \Eq{eq:c-M} relation for haloes belonging to voids (v) sheets (s) 
filaments(f) and knots (k).}
\label{tab:cm_env}
\begin{center}
\begin{tabular}{cccccc}
\hline
Parameter & \LCDM & \cde & \cde033 & \cde066 & \cde099 \\
\hline 
$c_0$(v)	 & $5.1 \pm0.5$   &  $5.36 \pm0.5 $&$  4.5 \pm0.2$ &  $ 4.5 \pm0.3$ & $ 4.3 \pm0.3 $ \\
$\gamma$(v) &$-0.03\pm0.01 $  & $-0.03\pm0.01 $& $-0.04\pm0.01$& $ -0.04\pm0.01$& $-0.03\pm0.01$\\
\hline
$c_0$(s)	 & $4.8 \pm0.1$   &  $5.1 \pm0.2 $&$  4.5 \pm0.1$ &  $ 4.3 \pm0.2$ & $ 4.2 \pm0.1 $ \\
$\gamma$(s) &$-0.034\pm0.007$  & $-0.037\pm0.009 $& $-0.04\pm0.01$& $ -0.04\pm0.01$& $-0.036\pm0.008$\\
\hline 
$c_0$(f)	 & $4.41 \pm0.05$   &  $4.42 \pm0.03 $& $  4.26 \pm0.05$ & $ 4.15 \pm0.04 $ & $ 3.98 \pm0.05 $ \\
$\gamma$(f) &$-0.052\pm0.01 $  & $-0.059\pm0.006 $& $-0.07\pm0.01$& $-0.064\pm0.008$& $-0.058\pm0.005$\\
\hline
$c_0$(k)	 & $4.06 \pm0.07$   &  $4.38 \pm0.08 $& $  4.02 \pm0.06$ & $ 3.74 \pm0.06 $ & $ 3.67 \pm0.07 $ \\
$\gamma$(k) &$-0.087\pm0.009$  & $-0.093\pm0.008 $& $-0.085\pm0.005$& $-0.093\pm0.009$& $-0.092\pm0.007$\\
\hline
\end{tabular}
\end{center}
\end{table*}

\subsection{Spin and concentration}
We now turn to the non-dimensional spin parameter, $\lambda$, and dark
matter halo concentrations, investigating how they will change across
different environments and cosmologies.  In the latter case, we will
also pay particular attention to the environment-related changes to
the $c-M$ relation of \Eq{eq:c-M}.  In both cases we refer to the
definitions introduced in \Sec{sec:halo_res}.

\subsubsection{Underdense regions: voids and sheets}
Looking at the median spin parameters shown in the upper panel of
\Fig{img:spin_env} we can again draw the conclusion that, just like in
the general case, \cde\ cosmologies lead to larger spins and that this
increase is proportional to the coupling parameter $\beta$. On the
other hand, the value of $\bar{\lambda}$ for haloes in \ude\
cosmologies is, on average, indistinguishable from \LCDM.  We can
therefore confirm the observation that underdense region contain
haloes with lower spins, just as found by \cite{Maccio:2007}.
However, the reduction in the median value is of the same order in all
models, so that combining the information of the environment does not
put tight constraints on the parameters of the model.

Concentrations, too, show a remarkable behaviour for haloes belonging
to underdense regions.  In \Tab{tab:cm_env} we show the results of
fitting the median concentration per mass bin to a power law,
i.e. \Eq{eq:c-M}.  The first thing we observe is that the correlation
between $c$ and $M$ (as measured by the power-law index $\gamma$) is
weaker than what we observed in the general case.
This, combined with the fact that in the lower mass bins
median concentration do not change with respect to the general case,
in turn leads to observed larger values for $c_0$, although the errors
are also large due to the small number statistics. However, some care
must be taken when considering this relation for void haloes since the
fit is based upon a small mass range only and also gives more weight
to lower mass objects \citep[][]{Prada:2012, DeBoni:2013}.

\subsubsection{Overdense regions: filaments and knots}
In the bottom panels of \Fig{img:spin_env} and \Fig{img:cm_env} we
plot the spin and concentration-mass relation; the best-fit values to
\Eq{eq:c-M} are again provided in \Tab{tab:cm_env}.

In the case of spins, we find that dark matter haloes in coupled
cosmologies tend to be characterized by larger values of $\lambda$.
However, haloes located in filamentary structures show, at least in
the lower mass bins, sharper differences between \cde\ models and
\LCDM\ than what is revealed by knots. This is also due to the smaller
number of low mass haloes living in knots, which visibly affects the
statistics of the parameter.

Concentrations instead show two slightly different patterns in
filaments and in knots.  In the former environment, all cosmologies
seem to be characterized by a flatter slope, which averages around
$-0.06$ and seems not to be connected to the underlying model.  In the
latter environment, a steeper correlation is found, with $\gamma
\approx -0.09$ -- much closer to the general case discussed in
\Sec{sec:halo_res}.  Not only the slope but also the normalization
$c_0$ of \Eq{eq:c-M} changes when considering filamentary or knot-like
environments: in the former case we find that this parameter is
substantially larger than in the latter.

Our results therefore indicate that the concentration-mass relation is not only
affected by the cosmological model but also by the environment the
haloes under consideration live in: $\gamma$ gets flatter while $c_0$
increases for decreasing densities.
However, at odds with what we found for halo number counts, we find here that environment
does not play a role in strengthening the magnitude of model-dependent properties
of haloes. While the effect of dark energy can still be clearly seen in the
higher spins and lower concentrations of dark matter haloes, these \cde-induced
characteristics are not enhanced by the environment.
In fact, whereas the halo content of the different regions depends on the 
model and reinforces the trends observed in \Sec{sec:halop}, the properties
of the haloes themselves, while still being correlated to the underlying cosmology,
seem to be shifted by the same amount as \LCDM.
This suggests that environmental effects, in these cases, influence to the same
extent both quintessential and standard models, and do not provide a stronger
model-specific kind of prediction.

\section{Conclusions}\label{sec:conclusions}
In the present work -- which forms part of a series of studies of
(coupled) dark energy models -- we have discussed the properties of
large-scale structures and the cosmic web as they emerge in a series
of different quintessence models, systematically comparing the results
of a coupled scalar field to those obtained for a free field and the
standard \LCDM\ cosmology.

We performed the following three-fold analysis:
\begin{itemize}
\item we studied halo mass function and general halo properties (mass, 
spin and concentrations),
\item we investigated the general properties of the cosmic web, using
  a kinetic classification algorithm,
\item we correlated halo properties to the environment.
\end{itemize}
First, we have studied several aspects of \cde\ and \ude\
cosmologies looking at the full halo sample. At this stage,
our results proved to be in line with those  of \cite{Baldi:2010a, Li:2011, Cui:2012},
finding that the analytical formulae for the halo mass functions
and dark matter profiles are valid also in this class of models.

Examining concentrations we found that, while \ude\ cosmology is
characterized by haloes with higher values for $c$, for \cde\ models
the opposite is in general the case -- in accordance with the results
of \cite{Baldi:2010a} and \cite{DeBoni:2013}.  
Interestingly, in the case of spin parameters
we observe a weak dependence on the coupling, since we can see that their
value is mildly enhanced by larger values of $\beta$, as was also noted by
\cite{Hellwing:2011} in the context of other fifth force cosmological
models.

The cosmic web investigated as part of this study is characterized by the
eigenvalues of the velocity shear tensor, a novel method recently
proposed by \cite{Hoffman:2012} and successfully applied to various
simulations by \cite{Libeskind:2012}.  
Computing the fraction of total mass and volume belonging to each type of environment in 
our cosmologies, we find that the structure of the cosmic web itself
does not reveal any particular difference among the models. 
The same conclusion can be drawn when investigating the global distribution 
of the shear tensor eigenvalues.

This notwithstanding, the classification of the cosmic web can be
extremely useful when married with the halo catalogue.
Combining the two, in fact, we were able to show that many of the
differences observed in some halo properties when studying a global
sample of relaxed structures above a threshold mass are in fact due to
objects belonging to a certain type of environment. This happens in particular
in voids and sheets, where the differences among \cde\ and \LCDM\ are
up to three times as large as they are in the general case.
We have been able to verify how the magnitude of this effect is closely
dependent on the coupling: while \cde\ cosmologies' underproduction of
haloes in these regions is largely amplified, the overproduction that
characterizes the \ude\ model investigated here is only weakly
enhanced.  This means that;
\begin{itemize}
 \item one should focus on voids and sheets (underdense regions) when
   looking for signatures of (coupled) dark energy, and
 \item the magnitude of the deviations from \LCDM\ allows us to place
   constraints on \cde\ cosmologies, or even detect them.
\end{itemize}

We have also seen how the standard concentration-mass relation is
substantially affected when fitted for halo samples classified
according to the environment they are located in.  While the standard
functional form of \Eq{eq:c-M} still holds in underdense regions for
all the models, it does so with a much steeper slope (the change is
from an average of $-0.9$ to $-0.4$) and a substantial increase in the
average concentration. In addition to this, we note again an amplification
of the difference between the $c_0$ values in \cde\ and \LCDM\ obtained when fitting
\Eq{eq:c-M} in voids and sheets (up to $15\%$) with respect to the
global one ($\approx7$ per cent).

The fact that these results are mostly visible when restricting the
halo sample to underdense environments tells us the importance of the
relative weights to be attached when performing global analyses.
Indeed, when referring to halo properties in general, we do in fact
hide a large number of peculiar features which can be seen only in a
narrower subset. In the particular case of \cde\ we have seen how
structures in voids play a major role, in the same direction of
\cite{Li:2011}, who also highlighted the importance of underdense
regions in the context of similar cosmological models.

As a concluding remark, we would like to emphasize that a great amount
of effects observed here still deserve a more in-depth study. In
particular, the analysis of the temporal evolution of halo parameters,
 will shed more light on
the mechanisms that result in the previously discussed differences at
$z=0$ and increase the observational features that can be used to
constrain quintessence models. We shall turn to these in future contributions
in this series.

\section*{Acknowledgements}
EC is supported by the {\it Spanish Ministerio de Econom\'ia y
  Competitividad} (MINECO) under grant no. AYA2012-31101, and
MultiDark Consolider project under grant CSD2009-00064. He further
thanks Georg Robbers for providing an updated, non-public version of
\texttt{CMBEASY}.

This work was undertaken as part of the Survey Simulation Pipeline (SSimPL: ssimpluniverse.tk)
and GFL acknowledges support from ARC/DP 130100117

AK is supported by the {\it Spanish Ministerio de Ciencia e
  Innovaci\'on} (MICINN) in Spain through the Ram\'{o}n y Cajal
programme as well as the grants AYA 2009-13875-C03-02, CSD2009-00064,
CAM S2009/ESP-1496 (from the ASTROMADRID network) and the {\it
  Ministerio de Econom\'ia y Competitividad} (MINECO) through grant
AYA2012-31101. He further thanks Emily for reflect on rye.

GY  acknowledges  support from  MINECO under research grants 
AYA2012-31101, FPA2012-34694,
Consolider Ingenio SyeC CSD2007-0050 and from Comunidad de Madrid under 
ASTROMADRID project (S2009/ESP-1496).

The authors  thankfully acknowledge the computer resources, technical expertise and assistance provided by the Red Espa\~nola de Supercomputaci\'on.

We further acknowledge partial support from the European Union FP7 ITN
INVISIBLES (Marie Curie Actions, PITN-GA-2011-289442).

All the simulations used in this work were performed in the
Marenostrum supercomputer at Barcelona Supercomputing Center (BSC).

\bibliographystyle{mn2e}
\bibliography{biblio}

\bsp

\label{lastpage}

\end{document}